\documentclass[prd,aps,a4,twocolumn,superscriptaddress,preprintnumbers,nofootinbib]{revtex4-1}
\usepackage{pslatex}
\usepackage[pdftex]{graphicx}
\usepackage{psfrag}
\usepackage{epsfig}
\usepackage[usenames,dvipsnames]{color}
\usepackage{cancel}
\usepackage{slashed}
\usepackage{amssymb}
\usepackage{amsmath}
\usepackage{hyperref}
\usepackage{enumerate}
\usepackage{multirow}
\usepackage{lipsum}

\definecolor{darkred}{rgb}{0.6,0,0}
\definecolor{dbrown}{rgb}{0.4,0.26,0.13}
\definecolor{linkcolor}{rgb}{0,0,0.5}

\begin{document}
\title{Non-standard cosmic expansion histories: Neutrino decoupling
  and primordial nucleosynthesis signatures}%
\author{D. Aristizabal Sierra}%
\email{daristizabal@ulg.ac.be}%
\affiliation{Universidad T\'ecnica
  Federico Santa Mar\'{i}a - Departamento de F\'{i}sica\\
  Casilla 110-V, Avda. Espa\~na 1680, Valpara\'{i}so, Chile}%
\author{S. Gariazzo}%
\email{stefano.gariazzo@ift.csic.es} \affiliation{Istituto Nazionale di Fisica
  Nucleare (INFN), Sezione di Torino, Via P. Giuria 1, I-10125 Turin,
  Italy}\affiliation{Instituto de Fisica Teorica, CSIC-UAM
C/ Nicolás Cabrera 13-15, Campus de Cantoblanco UAM, 28049 Madrid, Spain}%
\author{A. Villanueva}%
\email{angelodavg@gmail.com}%
\affiliation{Instituto de Física, Pontificia Universidad Catolica de Valparaiso (PUCV), 
Avda. Brasil 2950, Casilla 4059, Valparaiso, Chile}%
\begin{abstract}
  Cosmological scenarios with a non-standard equation of state can
  involve ultrastiff fluids, understood as primordial fluids for which
  $p/\rho> 1$. Their energy densities can dominate the Universe
  energy budget at early times, in the otherwise radiation dominated
  epoch. During that period the Universe undergoes a faster expansion,
  that has implications for any decoupling process that takes place in
  that era. Quintessence models or Ekpyrotic cosmologies are good
  examples of such scenarios. Assuming the ultrastiff state to be
  thermally decoupled at very early times, if ever coupled, its
  observational imprints are left solely in the Universe expansion
  rate and in the radiation energy density. We consider a complete set
  of ultrastiff fluids and study their signatures in the neutrino
  decoupling and BBN eras. Measurements of $N_\text{eff}$ alone place
  mild constraints on these scenarios, with forthcoming measurements
  from the Simons Observatory in the Chilean Atacama desert being able
  to test regions where still sizable effects are observable. However,
  when BBN data is taken into account, those regions are proven to be
  barely reconcilable with primordial helium-4 and deuterium
  abundances measurements. Our findings show that measurements of the
  primordial helium-4 abundance imply the tightest constraints, with
  measurements of primordial deuterium being---to a certain
  extent---competitive as well. We point out that a $\sim 60\%$
  improvement on the statistical uncertainty of the primordial
  helium-4 abundance measurement, will test these scenarios in the
  region where they can produce sizable effects. Beyond that precision
  the regions that are accessible degenerate with standard
  expectations. In that case, although potentially present, neither
  neutrino decoupling nor BBN observables will be sensitive probes.
\end{abstract}
\maketitle

\tableofcontents
\section{Introduction}
\label{sec:intro}
Measurements of light elements abundances along with CMB power spectra
provide the most early picture of the primordial Universe we have so
far. CMB power spectra strongly supports the $\Lambda\text{CDM}$
model, while at the same time placing tight constrains on possible new
effects at temperatures of the order of $1\;$eV
\cite{Planck:2018vyg}. Meanwhile Big Bang Nucleosynthesis (BBN),
triggered---roughly---when the average photon distribution energy
reaches values that allow light nuclei binding, embody the most
ancient prove of the physics that took place much before the formation
of the large-scale structure of the Universe. It furnishes a powerful
tool that enables testing physics beyond the $\Lambda\text{CDM}$ model
at temperatures of the order of 0.1 MeV \cite{Iocco:2008va}. In the
future, measurements of earlier epochs observables such as those from
the cosmic neutrino background or from gravitational waves will
provide a picture dating back to times where the primordial plasma was
at much higher temperatures
\cite{PTOLEMY:2019hkd,Punturo:2010zz,Yagi:2011wg,LIGOScientific:2014qfs,LIGOScientific:2016wof,LISA:2017pwj}.

Deviations from the $\Lambda\text{CDM}$ model can be induced by a
variety of new physics scenarios. However, given the excellent match
between observational data and $\Lambda\text{CDM}$ predictions, they
are expected to be small at least up to the BBN era. From the
microphysics point of view, quite generically, the deviations can be
categorized as being generated by: (i) New degrees of freedom
(particle physics), (ii) changes on the dynamics of the Universe
expansion rate (gravity, fluid composition), (iii) combinations of (i)
and (ii). This is somewhat expected. Densities of whatever kind are
tracked with kinetic equations which involve information on particle
and expansion dynamics. At the MeV scale imprints can be left in two
main observables, $N_\text{eff}$ (measured at either the BBN or CMB
era) and/or the relative abundance of light elements in particular
those of deuterium and helium-4. With $N_\text{eff}$ determined at the
ten-percent level
\cite{Planck:2018vyg,Fields:2019pfx,Brian_D_Fields_2020,Lisi:1999ng}
and the deuterium and helium-4 abundances at the percent level
\cite{10.1093/ptep/ptac097,Hsyu:2020uqb}, constraints arising from
early Universe physics are quite often even more stringent that those
derived from laboratory experiments (see
e.g.~\cite{Giovanetti:2021izc}).

A rather representative example of scenarios (i) is that of a light
thermal relic \footnote{MeV reheating scenarios are as well a rather representative example, see e.g.\ Refs.~\cite{Hasegawa:2019jsa,Kawasaki:2000en,Kawasaki:1999na}.}. Its presence in the early Universe contributes to the
radiation density and thus generates deviations on $N_\text{eff}$ and
on primordial abundances \cite{Giovanetti:2021izc,Lague:2019yvs}. In
category (ii) scenarios one could certainly mention scalar-tensor
theories \cite{Jordan1955-JORSUW,Fierz:1956zz,Brans:1961sx}, in
particular those realizations with an attractor mechanism towards
general relativity \cite{Bartolo:1999sq}. Their main phenomenological
effect is that of driving the Universe into a period of non-standard
expansion, which ultimately at the time of BBN converges towards that
predicted by the standard cosmological model.

An interesting possibility, beyond these two benchmark cases, is that
of an energy density following from a non-standard equation of
state. Possible theoretical scenarios that can lead to such
contribution include quintessence models
\cite{Ratra:1987rm,Wetterich:1987fm}, Ekpyrotic cosmologies
\cite{Khoury:2001wf,Khoury:2003rt} or models with a \textit{stiff
matter} component \cite{Zeldovich:1972zz,Chavanis:2014lra} \footnote{Note that although possible, construction of stiff fluid cosmological models is challenging \cite{Scherrer:2022nnz}.}. In the
latter case, the equation of state is given by $p_s=\rho_s$
($\omega_s=1$) resulting in an energy density component scaling as
$a^{-6}$. Consequently, there is a period during which---in the
otherwise radiation dominated epoch---the Universe becomes dominated
by the ultrastiff component (in this particular case called
\textit{stiff matter}). Since
$\rho_\text{rad}^\text{stand}<\rho_\text{rad}^\text{non-stand}$,
during that epoch the Universe is expected to expand at a faster rate,
thus leading to effects that involve dark matter freeze-out and
freeze-in \cite{DEramo:2017gpl,DEramo:2017ecx} and production of light
elements during the BBN epoch \cite{Dutta:2010cu}.

Studying non-standard expansion histories stands by its
own. Regardless of whether or not a certain equation of state fits
within a compelling theoretical framework, from a phenomenological
perspective, it is desirable to identify observational imprints that
such ``exotic'' contributions might have left in cosmological
observables.  The impact of such contribution in the radiation
dominated era is two-fold\footnote{Here we emphasize on imprints
  potentially left on observables already measured. There are of
  course other imprints one can think about. For instance, deviations
  on dark matter decoupling \cite{DEramo:2017gpl,DEramo:2017ecx} or
  effects on the generation of the cosmic baryon asymmetry through
  leptogenesis. The latter has been already considered, but in a
  different class of non-standard expansion history scenarios
  (scalar-tensor theories) \cite{Dutta:2018zkg}.}: First of all, the
energy density of such background will unavoidably contribute to the
relativistic degrees of freedom $N_\text{eff}$, whose value is fixed
at the time of neutrino decoupling. This calls for an analysis of
neutrino decoupling dynamics in the presence of such new energy
density. Secondly, since the expansion is affected so will the
formation of light elements abundances. This, in turn, calls for an
analysis of BBN dynamics. This is what this paper aims at.

As far as we know, implications of stiff matter on the generation of
light elements during the BBN era was first investigated in
Ref.~\cite{Dutta:2010cu}. Implications of a stiff fluid---as well as
of other kind of components---for dark matter freeze-out and freeze-in
have been comprehensively considered in
Refs.~\cite{DEramo:2017gpl,DEramo:2017ecx}. Using $N_\text{eff}$ at
the time of BBN (taken to be 1 MeV), this reference has determined as
well the constraints that these type of scenarios should obey. More
recently, constraints from BBN as well as from CMB have been derived
for axion kination in Ref. \cite{Co:2021lkc}. Here we extend upon the
analyses of these references by: (i) Considering the full process of
neutrino decoupling with the corresponding network of kinetic
equations, (ii) studying BBN with state-of-the-art numerical tools in
cases beyond the stiff matter scenario.

Considering neutrino decoupling seems mandatory. With a faster
expansion rate the evolution of the photon and neutrino densities
(temperatures) are expected to change. The exotic component will have,
therefore, a two-fold effect on $N_\text{eff}$: Through its change on
$T_\gamma/T_\nu$ and through its direct contribution to the radiation
density. The neutrino decoupling process is also relevant since it
determines the temperature at which entropy release from $e^+e^-$ pair
annihilation ceases, required for a proper evaluation of
$N_\text{eff}$. For the same reason that neutrino decoupling should
be considered, a precise determination of constraints in parameter
space implied by BBN calls for a detailed tracking of light elements
densities. A faster expansion rate is expected to accelerate
electroweak processes decoupling. The neutron-to-proton ratio is then
expected to departure from its standard expectation, thus
affecting---potentially sizably---the $^4\text{He}$ abundance along
with the other abundances.

The remainder of this paper is organized as follows. In
Sec.~\ref{sec:nonstandard-cosmo}, we briefly discuss our notation and
the parametrization that allows a model-independent treatment of the
ultrastiff components. In
Sec.~\ref{sec:constraints_nu_decoupling_CMB}, after discussing the
current status of measurements of $N_\text{eff}$ and future
improvements, we present our analysis of neutrino decoupling in these
kind of scenarios. Sec.~\ref{sec:constraints_BBN} discusses the
generation of the light elements abundances, along with a short review
of the status of the measurements of primordial deuterium and helium-4
abundances. Finally, in Sec.~\ref{sec:conclusions} we conclude and
present our summary. In App.~\ref{sec:thermodynamics} and
\ref{sec:proton_neutron_rates} a few relevant standard results are
given.

\section{Parametrization of ultrastiff components and non-standard
  expansion rate}
\label{sec:nonstandard-cosmo}
Phenomenological consequences of ultrastiff fluids can be derived by
using a simple---yet powerful---model-independent parametrization
first introduced, as far as we know, in
Ref.~\cite{DEramo:2017gpl}. Since our analyses rely heavily on that
parametrization here we briefly review its main aspects. Pressure and
energy density can always be related through the simple equation of
state $p=\omega \rho$. For a time-independent proportionality
constant, integration of the fluid equation leads to
$\rho=\rho_0\,a^{-3(1+\omega)}$ ($\rho_0=\text{const}$). Matter,
radiation and vacuum energy densities are characterized by
$\omega_\text{mat}=0$, $\omega_\text{rad}=1/3$ and
$\omega_\Lambda=-1$. As previously pointed out, the case of stiff
matter is defined by $\omega=1$. Barring the case $\omega=2/3$, all the
other scenarios are defined by $\omega>1$, which include the case of
Ekpyrotic cosmologies. Note that ultrastiff fluids are defined by the
condition $p/\rho>1$ (see e.g.~\cite{Barrow:2010rx}), we
include---however---$\omega=2/3$ and $\omega=1$ in that definition.

In order to compare with radiation it proves useful to introduce
\begin{equation}
  \label{eq:rho_s_param}
  \rho_s(a)=\rho(a_c)\left(\frac{a_c}{a}\right)^{4+n}\ ,
\end{equation}
where $a_c$ is the value of the scale factor at crossover time, when
$\rho_\text{rad}$ and $\rho_s$ match. From now on we will refer to
that epoch as the \textit{ultrastiff-radiation equality
  era}. Departures from standard radiation, as those we are interested
in, require $n>1$ with $\omega=(1+n)/3$. Since we are interested in
tracking densities with kinetic equations it proves convenient to
rewrite Eq.~(\ref{eq:rho_s_param}) in terms of
temperature. Conservation of entropy per comoving volume,
$g_{\star S}(T)\,a^3\,T^3=\text{const}$, enables that. The result
reads
\begin{equation}
  \label{eq:rho_stiff}
  \rho_s(T_\gamma,T_c,n)=\rho_s(T_c)
  \left[
    \frac{g_{\star S}(T_\gamma)}{g_{\star S}(T_c)}
  \right]^{(4+n)/3}
  \left(\frac{T_\gamma}{T_c}\right)^{4+n}\ .
\end{equation}
Here $T_c$ refers to the temperature of the heat bath at
ultrastiff-radiation equality and $n$ determines the stiffness of the
ultrastiff fluid. Note that in the temperature range we are interested
in, $T_\gamma\subset [10^{-2},10]\,\text{MeV}$,
$g_{\star S}(T_\gamma)$ has a temperature dependence determined by the
photon overheating implied by electron-positron annihilation after
neutrino decoupling. Note that $\rho_s(T_c)=\rho_\text{rad}(T_c)$,
where $\rho_\text{rad}(T_c)$ is the standard expression for radiation
\begin{equation}
  \label{eq:rad_density}
  \rho_\text{rad}(T_\gamma)=\frac{\pi^2}{30}g_\star(T_\gamma)
  T_\gamma^4\ ,
\end{equation}
evaluated at the crossover temperature. The energy budget thus follows
from
\begin{equation}
  \label{eq:energy_budget}
  \rho_\text{Tot}(T_\gamma,T_c,n)=\rho_\text{rad}(T_\gamma)
  +\rho_s(T_\gamma,T_c,n)\ .
\end{equation}
The amount of energy density in the heat bath is now a
parameter-dependent quantity and so the expansion rate
\begin{equation}
  \label{eq:expansion_rate}
  H(T_\gamma,T_c,n)=\sqrt{\frac{8\pi}{3m_\text{Pl}^2}
    \rho_\text{Tot}(T_\gamma,T_c,n)}\ .
\end{equation}
From Eqs.~(\ref{eq:energy_budget}) and (\ref{eq:expansion_rate}) it
becomes clear that imprints of the ultrastiff fluid are left in the
amount of extra radiation as well as in the expansion rate. The latter
has---of course---ramifications in any decoupling process occurring in
the heat bath at the time when the contribution of the ultrastiff
fluid is still relevant.

In the setups we are considering the ultrastiff fluid is assumed to
have gone through decoupling, if ever thermally coupled, at very early
times. Thus, it has no implications whatsoever in the microphysical
processes taking place in the heat bath. The evolution of $\rho_s$ is
dictated by a collisionless Boltzmann equation. The collision terms in
the network of Boltzmann equations we employ in the next sections are
thus unaffected by the presence of the ultrastiff fluid.

\section{Neutrino decoupling}
\label{sec:constraints_nu_decoupling_CMB}
During the neutrino decoupling era the presence of an ultrastiff fluid is
expected to affect, potentially sizably, $N_\text{eff}$ (the amount of
radiation beyond photons). The most obvious way is by its direct
contribution to the bulk of the radiation energy density. The none so
obvious way, however, is through the impact on the expansion
rate. Modification of the expansion rate at this epoch is expected to
have an impact on the evolution of both the neutrino and photon
temperatures, $T_\nu$ and $T\equiv T_\gamma$.

A simplified picture of neutrino decoupling in the standard model is
as follows. At temperatures of the order of $10\,$MeV, species are in
thermal contact through electroweak interactions \footnote{Through
  electromagnetic interactions as well, but these are irrelevant for
  neutrino decoupling.}. Because of the low temperature, the most
abundant species are $e^\pm$ pairs, photons and the three neutrino
flavors, which at this epoch are still relativistic. Densities for all
the other species are Boltzmann suppressed and so have a negligible
effect (the exception being nucleons which, despite their small
densities, become relevant afterwards as their densities feed the
formation of the light elements at $T\simeq 0.1\,$MeV, see
Sec.~\ref{sec:constraints_BBN}). As the temperature decreases,
neutrino-electron electroweak scattering processes slow down (compared
with the expansion rate), and so at $T\simeq 1\,$MeV neutrinos
decouple from the thermal bath. Soon after neutrino decoupling---at
$T\lesssim m_e$---electron-positron pair production becomes
inefficient (because of kinematic reasons), thus favoring entropy
injection to the hot plasma. The photon thermal distribution gets
overheated, but not the neutrino distributions which are already
decoupled. The number of relativistic degrees of freedom during this
epoch, along with entropy conservation leads to $T_\nu/T=(4/11)^{1/3}$
(see \cite{Lesgourgues:2006nd} for a detailed discussion).

In the standard model, during the radiation dominated epoch, neutrinos
fix the expansion rate. The amount of radiation at the end of the
neutrino decoupling era can be parameterized according to
\cite{Shvartsman:1969mm,Steigman:1977kc}
\begin{equation}
  \label{eq:rho_r}
  \rho_\text{rad}=
  \left[
    1 + \frac{7}{8}\left(\frac{4}{11}\right)^{4/3}N_\text{eff}
  \right]\,\rho_\gamma\ ,
\end{equation}
from which an expression for $N_\text{eff}$ follows
\begin{equation}
  \label{eq:Neff}
  N_\text{eff} = \frac{8}{7}\left(\frac{11}{4}\right)^{4/3}
  \left(
    \frac{\rho_\text{rad} - \rho_\gamma}{\rho_\gamma}
  \right)\ .
\end{equation}
In this way $N_\text{eff}$ parametrizes not only the neutrino
contribution to the radiation energy density, but all other radiation
components, if any. Assuming that only neutrinos and photons are
present and that their temperatures suffice to specify their
distributions, the leading order standard model expectation is
$N_\text{eff}=3$. Subleading effects including QED thermal
corrections, neutrino flavor oscillations and a full assessment of the
neutrino-neutrino collision integral have been studied in
Refs.~\cite{Bennett:2019ewm,Bennett:2020zkv,Froustey:2020mcq,Froustey:2021azz}. With these corrections
accounted for the standard model value is given by
\cite{Bennett:2020zkv} (see also Refs.~\cite{Akita:2020szl,Froustey:2020mcq})
\begin{equation}
  \label{eq:Neff_SM}
  N_\text{eff}^\text{SM} = 3.0440 \pm 0.0002\ ,
\end{equation}
where uncertainties are dominated by numerics and measurements of the
solar neutrino mixing angle. Results from a more recent analysis in
Ref.~\cite{Cielo:2023bqp} are inline with those in
Eq.~(\ref{eq:Neff_SM}).

Observationally, values for $N_\text{eff}$ follow either from CMB or
BBN data sets. Using Planck satellite data, fits of the standard
cosmological model lead to \cite{Planck:2018vyg}
\begin{equation}
  \label{eq:Neff_LambdaCDM}
  N_\text{eff}^\text{CMB} = 2.99\pm 0.17\quad (68\%\;\text{CL})\ ,
\end{equation}
whereas values inferred from primordial nucleosynthesis are instead
given by \cite{Consiglio:2017pot,Fields:2019pfx,Brian_D_Fields_2020,Lisi:1999ng}
\begin{equation}
  \label{eq:Neff_BBN}
  N_\text{eff}^\text{BBN} = 2.843\pm 0.154\quad (95\%\;\text{CL})\ .
\end{equation}
Improvements upon these values are expected from the Simons
Observatory in the Chilean Atacama desert (stage-III CMB experiment)
\cite{SimonsObservatory:2018koc}, which aims at a precision of
$\sigma(N_\text{eff})=0.05$. Further improvements are expected as well
from CMB-S4 \cite{Abazajian:2019eic}, a follow-up stage-IV CMB
experiment, that aims at a precision of $\sigma(N_\text{eff})=0.03$~\cite{CMB-S4:2016ple}.
\begin{figure*}[t]
  \centering
  \includegraphics[scale=0.382]{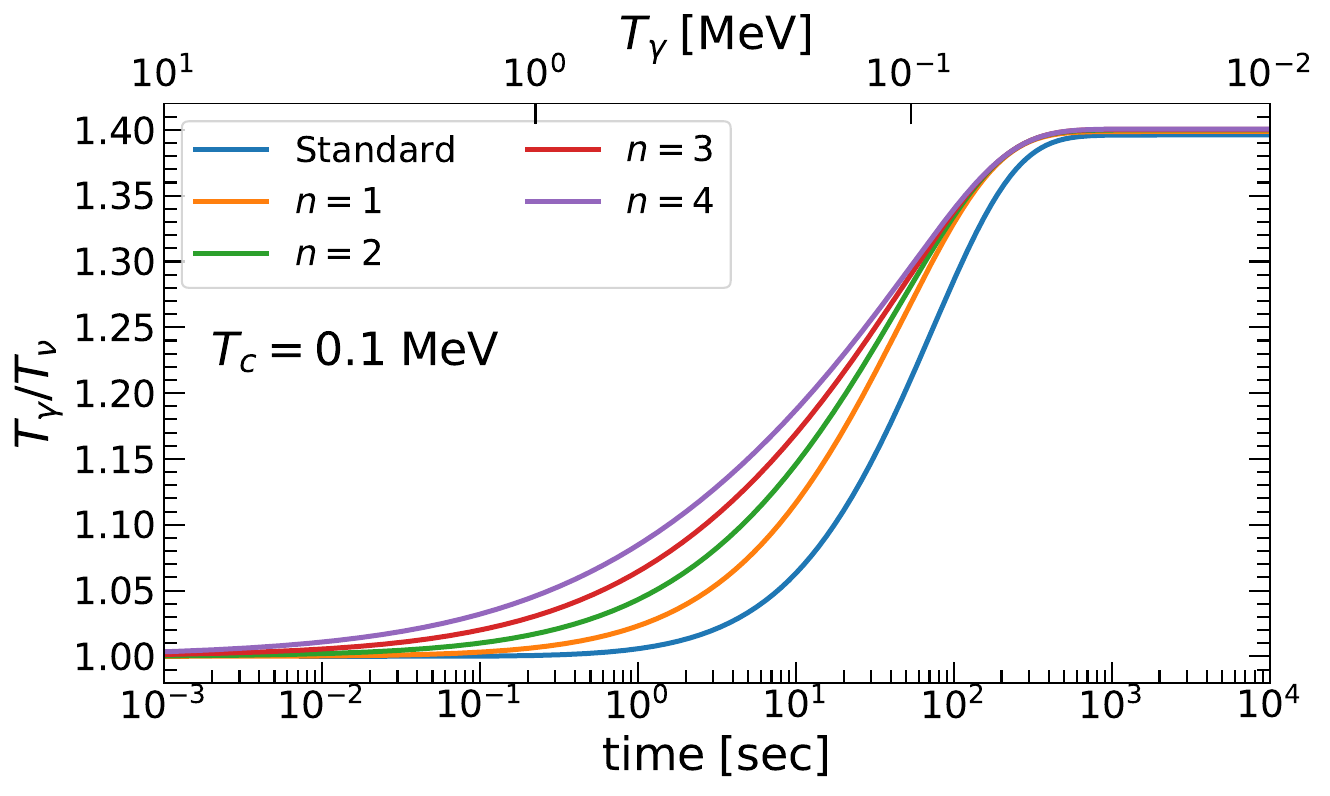}
  \includegraphics[scale=0.382]{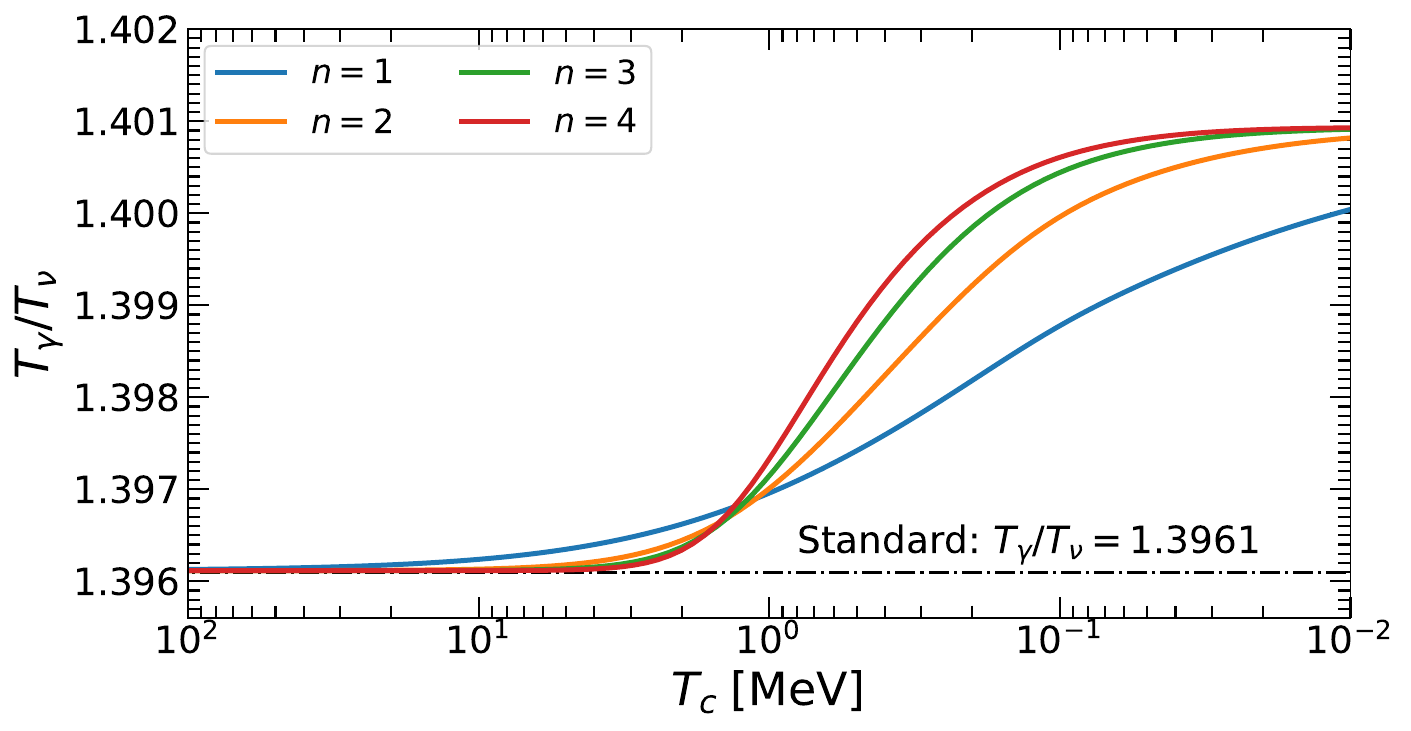}
  \caption{\textbf{Left graph}: $T_\gamma/T_\nu$ ratio as a function
    of time for different ultrastiff fluid scenarios $n=1,\cdots, 4$ and
    for a crossover temperature $T_c=0.1\,$MeV. The result has been
    derived by integration of the set of Boltzmann equations in
    Eqs.~(\ref{eqs:Beqs_nu}) and (\ref{eqs:Beqs_gam}). Because of the
    faster expansion rate, $e^+e^-$ pair annihilation injects entropy
    to the thermal bath at earlier times. Overheating of the photon
    distribution temperature changes at the permille level above the
    standard model expectation. The upper axis, shown just as a guide,
    displays the temperature calculated assuming $t=1/2/H$. An
    accurate expression, used for integration, follows from
    Eq.~(\ref{eq:jacobian}). \textbf{Right graph}: $T_\gamma/T_\nu$
    versus crossover temperature $T_c$ for the ultrastiff fluid cases
    $n=1,\cdots, 4$ (see text for further details). Shown as well is
    the standard model value without QED finite temperature
    corrections \cite{Escudero:2018mvt}.}
  \label{fig:tg_tnu_ratio}
\end{figure*}

\subsection{Neutrino decoupling in an ultrastiff dominated
  cosmological era}
\label{sec:nu_dec_stiff_cosmo_era}
With the ultrastiff contribution and assuming an universal temperature
for the three neutrino flavor species ($T_{\nu_i}=T_\nu$),
Eqs.~(\ref{eq:rho_stiff}) and (\ref{eq:Neff}) allow writing
\begin{widetext}
\begin{equation}
  \label{N_eff_stiff}
  N_\text{eff} = \left(\frac{11}{4}\right)^{4/3}
  \left\{3
  \left(\frac{T_\nu}{T}\right)^4
  + \frac{4\;g_\star(T_c)}{7}
  \left[\frac{g_{\star S}(T)}{g_{\star S}(T_c)}\right]^{(4+n)/3}
  \left(\frac{T}{T_c}\right)^n
  \right\}\ .
\end{equation}
\end{widetext}
As expected, this result converges to the standard case in the absence
of the ultrastiff contribution component. From this expression it becomes
clear that variations on $N_\text{eff}$ can be due to variations on
$T$ and $T_\nu$ (implied by a faster expansion rate), by the ultrastiff
component itself or more generally by both. To determine the extent at
which the photon and neutrino temperature changes because of a faster
expansion rate we rely on a simplified, yet precise, set of Boltzmann
equations: Relevant species are assumed to follow thermal equilibrium
distributions, thermally-averaged cross sections assume
Maxwell-Boltzmann distributions, the electron mass in the collision
terms as well as chemical potentials are neglected. The simplified
system of Boltzmann equations reads \cite{Escudero:2018mvt,Escudero:2020dfa}
\begin{align}
  \label{eqs:Beqs_nu}
  \frac{dT_\nu}{dt}&=-HT_\nu
                     +
                     \frac{\frac{\delta\rho_{\nu_e}}{\delta t}
                     + 2\frac{\delta\rho_{\nu_\mu}}{\delta t}}{3\frac{\partial\rho_\nu}
                     {\partial T_\nu}}\ ,
  \\
  \label{eqs:Beqs_gam}
  \frac{dT}{dt}&=
                        -\frac{
                        4H\rho_\gamma + 3H(\rho_e+p_e)+\frac{\delta\rho_{\nu_e}}{\delta t}
                        +
                        2\frac{\delta\rho_{\nu_\mu}}{\delta t}}{\frac{\partial\rho_\gamma}
                        {\partial T}+ \frac{\partial\rho_e}{\partial T}
                        }\ .
\end{align}
Here $\delta\rho_{\nu_e}/\delta t$ and $\delta\rho_{\nu_\mu}/\delta t$
refer to collision terms. Their explicit expressions along with
expressions for the other relevant thermodynamical quantities are
given in Appendix \ref{sec:thermodynamics}. Note that the results
derived with Eqs.~(\ref{eqs:Beqs_nu}) and (\ref{eqs:Beqs_gam}),
presented below, have been double-checked with the aid of the
\texttt{FortEPiaNO} public code
\cite{Gariazzo:2019gyi,Bennett:2020zkv}. Results in both cases are
inline.

We start with the determination of departures of $T/T_\nu$ from the
standard expectation, $T/T_\nu=1.3961$
\cite{Escudero:2018mvt}. Considering (as a benchmark scenario) the
case $T_c=0.1\,$MeV for four different ultrastiff fluid choices
$n=1,\cdots,4$, we calculated the time evolution of the neutrino and
photon temperatures (densities) with the aid of
Eqs.~(\ref{eqs:Beqs_nu}) and (\ref{eqs:Beqs_gam}). For the
determination of the time integration limits we have used the
time-temperature Jacobian
\begin{equation}
  \label{eq:jacobian}
  \left|\frac{dt}{dT}\right|=\frac{1}{H(T,T_c,n)\,T}
  \left[
    1 + \frac{T}{3g_{\star S}(T)}\frac{d}{dT}g_{\star S}(T)
  \right]\ ,
\end{equation}
which follows from entropy conservation per comoving volume and which
reduces to the standard result $t=1/2/H(T)$ in the absence of $\rho_s$
and during an epoch of no entropy transfer. Note that the relation
between time and temperature is affected because of the ultrastiff
contribution. Results are shown in the left graph in
Fig.~\ref{fig:tg_tnu_ratio}.

With increasing $n$ (increasing stiffness), departure from $T_\nu=T$
starts at earlier times (higher $T$). A stiffer energy density
component implies a faster expansion rate which, in turn, implies
processes to be less effective than otherwise (they abandon their
thermal behavior at earlier times). Thus, while at sufficiently high
temperatures $e^+e^-$ pair production ($\gamma+\gamma\to e^++e^-$) is
as likely as annihilation ($e^++e^- \to \gamma+\gamma$), production
ceases to be effective at earlier times because of the higher
expansion rate. Entropy starts being released in the hot plasma at
earlier times and so the earlier the departure from $T=T_\nu$. The
trend is therefore as follows. The stiffer the component, the more
overheated the photon distribution is at the end of the $e^+e^-$ pair
annihilation era. For the benchmark scenario we are considering here
($T_c=0.1\;\text{MeV}$) the largest departure is found to be
\footnote{For the percentage relative change in $T/T_\nu$ we
  use the notation $\Delta_i^{\gamma\nu}$, where $i$ refers to $n_i$
  and the difference is calculated with respect to the standard case.}
\begin{equation}
  \label{eq:departures_from_std_Tg_Tnu}
  \Delta_4^{\gamma\nu}=\frac{T/T_\nu|_{n=4}- T/T_\nu|_\text{Stand}}
  {T/T_\nu|_{n=4}}\times 100\%=0.3\%\ .
\end{equation}

To determine the general behavior we have calculated $T/T_\nu$ as a
function of $T_c$. Results are shown in the right graph in
Fig.~\ref{fig:tg_tnu_ratio} for the ultrastiff fluid scenarios
$n=1,\cdots, 4$. First of all, one can see that at high $T_c$ values
scenarios with small $n$ (less stiff) tend to deviate more from the
standard expectation. This behavior can be readily understood from
Eq.~(\ref{eq:rho_stiff}) coupled with results in the left graph of the
same figure. From that graph it is clear that $e^+e^-$ pair
annihilation ceases at $T\simeq 5\times 10^{-2}\;$MeV, with small
variations about that number depending on $n$ \footnote{Recently a
  detailed calculation of $e^+e^-$ pair annihilation freeze-out,
  including Fermi blocking and Bose enhancement as well as chemical
  potentials, has shown that $e^+e^-$ pair annihilation ceases at
  $T\simeq 16\,$keV \cite{Thomas:2019ran}.}. Values shown in the right
graph in Fig.~\ref{fig:tg_tnu_ratio} are those at that value, although
integration over time covers values down to $T\simeq 10^{-2}\,$MeV. At
large values of $T_c$, the suppression already implied by a small
value of $T$ is further enhanced with increasing values of $n$. So,
the smaller $n$ the larger the deviation in $H$ implied by
$\rho_s$. Conversely, for small values of $T_c$ (comparable to the
temperature where $e^+e^-$ pair annihilation ceases) larger values of
$n$ produce the largest deviations inline with the result displayed in
the graph.

The largest deviations on $T/T_\nu$ are found for small $T_c$ and
large $n$, a result somewhat expected. It is for those configurations
in parameter space for which one finds the largest deviations on the
expansion rate, compared with the standard expansion. For the cases we
have considered and up to $T_c=10^{-2}\;$MeV, deviations from the
standard case are all at the permille level:
\begin{align}
  \label{eq:deviations_Tg_Tn_freezeout}
  \Delta_1^{\gamma\nu}=&0.280\%\ ,\quad \Delta_2^{\gamma\nu}=0.336\%\ ,
  \\
  \label{eq:deviations_Tg_Tn_freezeout_1}
  \Delta_3^{\gamma\nu}=&0.342\%\ ,\quad \Delta_4^{\gamma\nu}=0.344\%\ .
\end{align}
Increasing $n$ and/or further decreasing $T_c$ will lead to larger
deviations. However, those configurations in parameter space
are---potentially---not reconcilable with constraints from light
elements abundances, in particular with those from $^{4}\text{He}$
(see Sec.~\ref{sec:constraints_BBN}). Note that if traces of the
ultrastiff fluid were determined only by these deviations,
measurements of $N_\text{eff}$ at the same level of those in
Eqs.~(\ref{eq:deviations_Tg_Tn_freezeout}) and
(\ref{eq:deviations_Tg_Tn_freezeout_1}) would be required to test
these scenarios. However, in addition to these deviations, there is as
well the contribution of the ultrastiff fluid energy density to
$N_\text{eff}$ that we now discuss.

\begin{figure*}[t]
  \centering
  \includegraphics[scale=0.405]{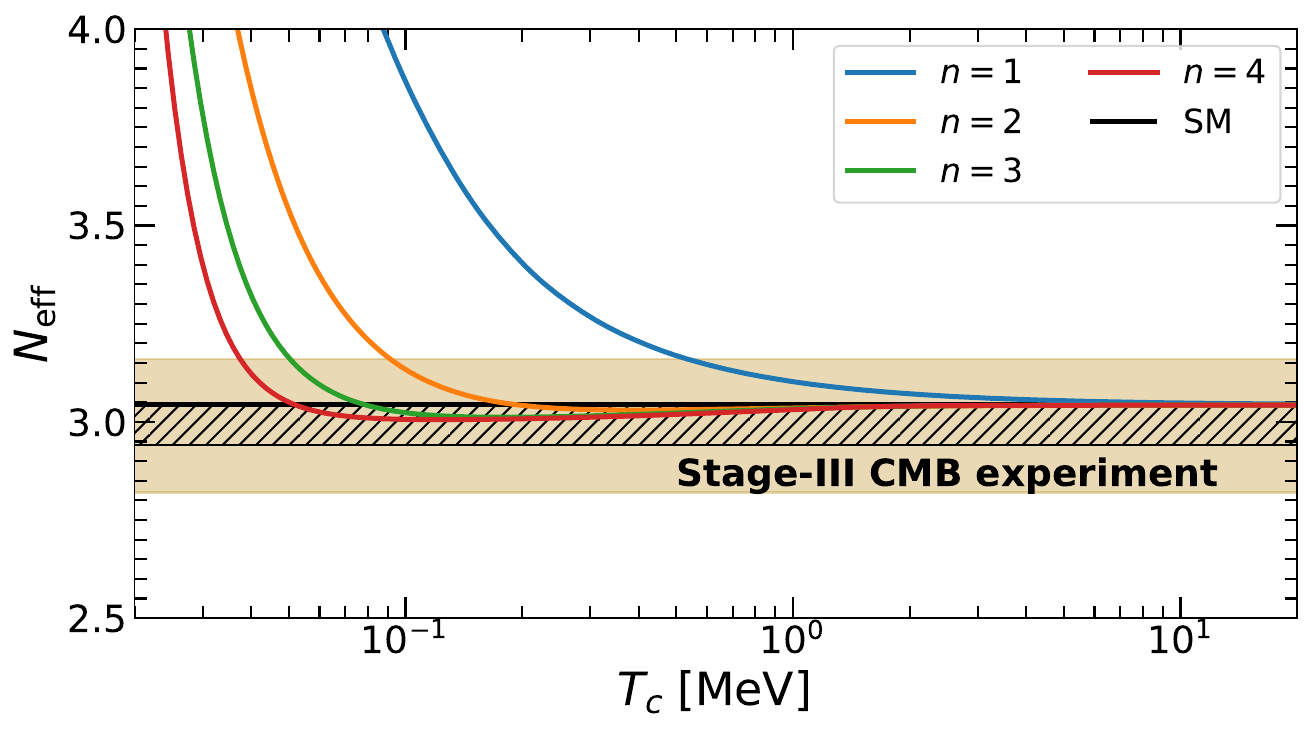}
  \includegraphics[scale=0.405]{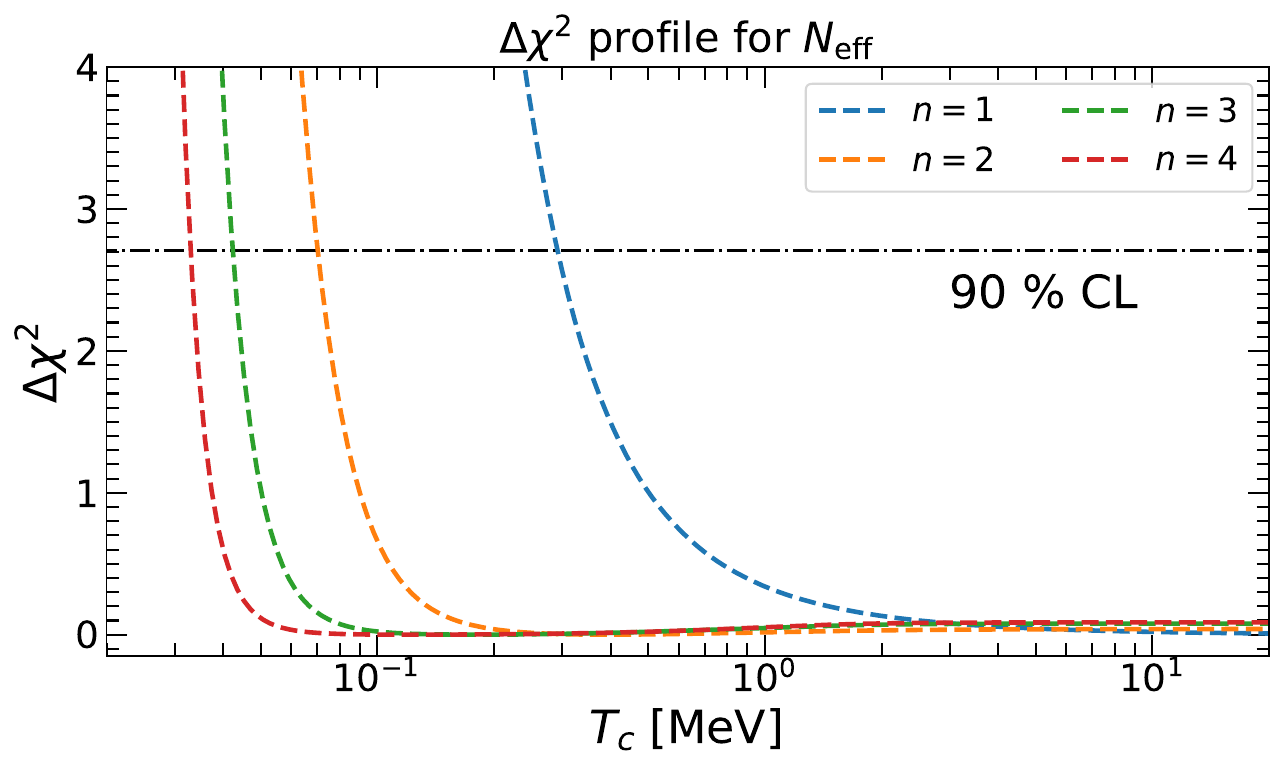}
  \caption{\textbf{Left graph}:$N_\text{eff}$ as a function of $T_c$
    for $n=1,\cdots,4$. The colored wide stripe corresponds to
    $N_\text{eff}^\text{CMB}$ as inferred from Planck data
    \cite{Planck:2018vyg} and given in
    Eq.~(\ref{eq:Neff_LambdaCDM}). The hatched region, instead, shows
    forecast values for $N_\text{eff}$, as expected from the Simons
    Observatory in the Chilean Atacama desert
    \cite{SimonsObservatory:2018koc}. \textbf{Right graph}:
    $\Delta\chi^2$ profile for the same cases shown in the left
    graph. The $90\%$ CL upper limits on $T_c$ are shown in
    Tab.~\ref{tab:Tc_Neff}.}
  \label{fig:N_eff}
\end{figure*}

To determine the effect of $\rho_s$ on $N_\text{eff}$ we have
calculated the second term in Eq.~(\ref{N_eff_stiff}) along with the
values for $T/T_\nu$ determined by Eqs.~(\ref{eqs:Beqs_nu}) and
(\ref{eqs:Beqs_gam}). Results for $N_\text{eff}$ as a function of
$T_c$ are displayed in the left graph in Fig.~\ref{fig:N_eff}. The
behavior is as follows. $N_\text{eff}$ converges to its standard value
at high $T_c$. This is somewhat expected, high values of $T_c$
suppress the contribution of $\rho_s$ to $N_\text{eff}$ and produce an
expansion that resembles that of the $\Lambda\text{CDM}$ model. As
$T_c$ decreases, both the contribution from $\rho_s$ and the
deviations on $T/T_\nu$ implied by a faster expansion rate kick in,
and so $N_\text{eff}$ deviates from the standard expectation. The less
stiff the contribution to $N_\text{eff}$ is, the larger the value of
$T_c$ where the departure starts growing rapidly. This behavior can be
seen in the graph, where sizable deviations are found (as $T_c$
decreases) first for $n=1$ then for $n=2$ and so on and so forth.

Using the current $68\%$ CL values for $N_\text{eff}^\text{CMB}$ in
Eq.~(\ref{eq:Neff_LambdaCDM}) one finds that consistency with current
measurements
require---roughly---$T_c\gtrsim \{50.0,9.00,5.00,2.00\}\times
10^{-2}\,$MeV for $n=\{1,2,3,4\}$. As can be seen in the graph,
improvements on $N_\text{eff}$ measurements---as those expected in
stage-III CMB experiments---will be able to explore regions in
parameter space that so far have not been
tested\footnote{$N_\text{eff}$ has been calculated at the temperature
  at which electron-positron annihilation ceases, which to a large
  extent amounts to the temperature at which the formation of the
  light elements abundances freeze-out. For our analysis we have
  employed, however, $N_\text{eff}^\text{CMB}$ which matches
  $N_\text{eff}^\text{BBN}$ within error bars. Note that here we are
  assuming that this will be the situation as well with stage-III CMB
  experiments.}. However, these results have to be coupled with those
following from BBN. And, as it will be shown in
Sec.~\ref{sec:constraints_BBN}, those regions are already ruled out by
primordial deuterium and helium-4 abundances measurements. Thus, if
any deviation from expectation is found in these round of experiments
that will further rule out regions of parameter space.

To put these results on a more statistical basis we have
performed a $\chi^2$ analysis with the aid of the following
least-square function
\begin{equation}
  \label{eq:chi2_Neff}
  \chi^2_\text{CMB}(T_c,n)=
  \left[
    \frac{N_\text{eff}^\text{Th}(T_c,n) - N_\text{eff}^\text{CMB}}
    {\sigma(N_\text{eff})}
  \right]^2\ ,
\end{equation}
where for $\sigma(N_\text{eff})$ we employ the statistical uncertainty
in Eq.~(\ref{eq:Neff_LambdaCDM}). Results are presented in the right
graph in Fig.~\ref{fig:N_eff}. Values for $T_c^\text{min}$ can be read
directly from the graph and are given in Tab.~\ref{tab:Tc_Neff} as a
reference. As has been already stressed, these results need to be
compared with those derived from the BBN analysis discussed in the
following section.
\begin{table}[h]
  \centering
  \begin{tabular}{|c|c|c|c|c|}\hline
    $n$ & 1 & 2 & 3 & 4\\\hline\hline
    $T_c^\text{min}\times 10^{-2}\;$MeV & 29.1 & 6.98 & 4.23 & 3.29\\\hline
  \end{tabular}
  \caption{$90\%$ CL lower limit for $T_c$ and for cases
    $n=1,\cdots,4$. The results follow after calculating
    $N_\text{eff}$ with the aid of Eqs.~(\ref{N_eff_stiff}),
    (\ref{eqs:Beqs_nu}) and (\ref{eqs:Beqs_gam}) and using the
    $\chi^2$ function given in Eq.~(\ref{eq:chi2_Neff}).}
  \label{tab:Tc_Neff}
\end{table}
\section{Primordial nucleosynthesis signatures}
\label{sec:constraints_BBN}
\begin{figure*}
  \centering
  \includegraphics[scale=0.45]{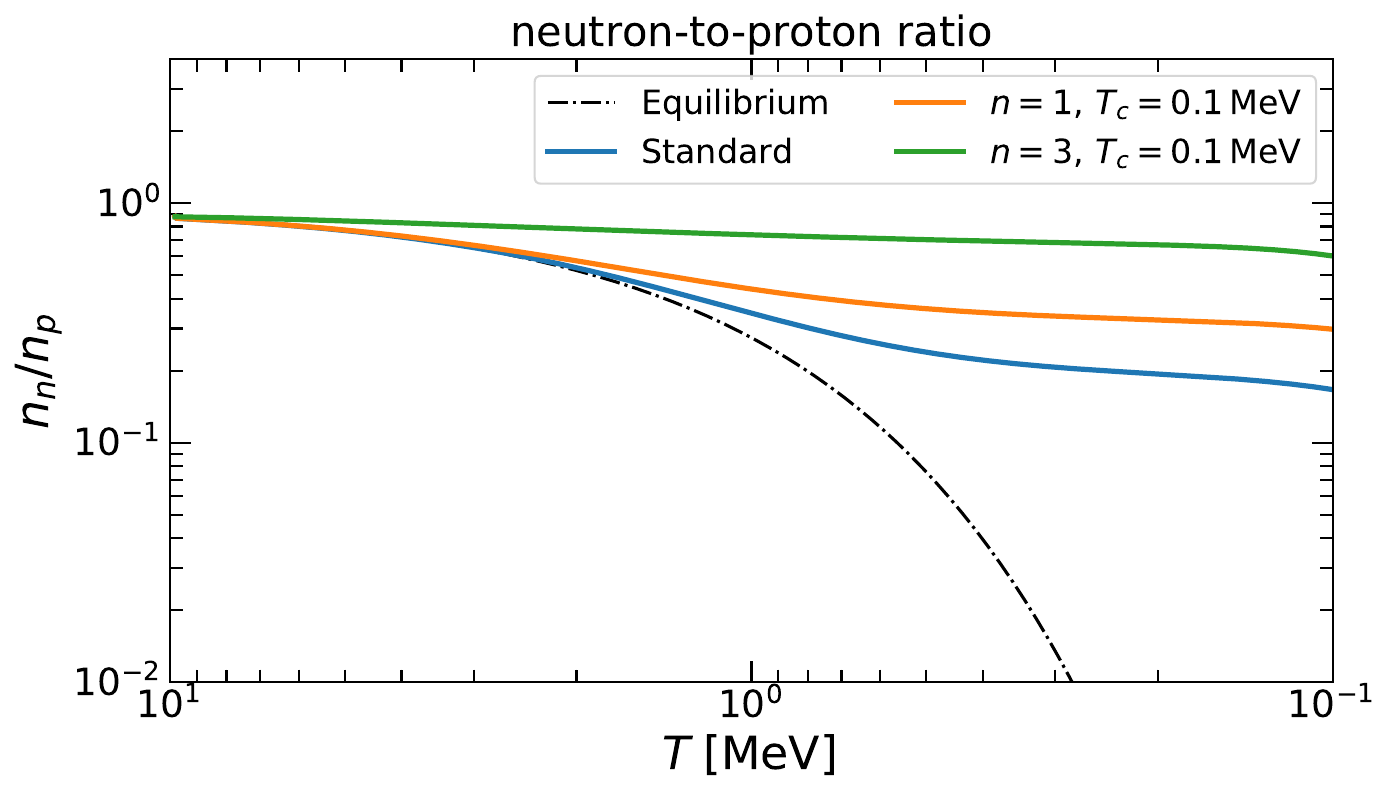}
  \caption{Neutron-to-proton ratio as a function of temperature in the
    standard case as well as in the non-standard cases $n=1$ and $n=3$
    for $T_c=0.1\,$MeV. For comparison, the thermal distribution is
    also shown. The stiffer the contribution to the energy density is,
    the higher the temperature at which the neutron-to-proton ratio
    deviates from its thermal equilibrium distribution. This behavior
    has an impact on BBN predictions.}
  \label{fig:deuterium_plots}
\end{figure*}
In this section we determine the impact of ultrastiff fluids on the
formation of the light elements abundances. We first start in
Sec.~\ref{sec:standard_case} with the standard scenario, aiming at a
description of the main qualitative features that the generation of
the deuterium ($d$), tritium ($t$), helium-3 and helium-4 abundances
obey. Note that a faster expansion rate is expected---in general---to
overproduce all the abundances. This can be readily and qualitatively
understood as follows. The neutron-to-proton ratio is expected to
deviate from its thermal distribution at higher $T$. The amount of
neutrons available at the onset of BBN is therefore larger and so---in
principle---all the light elements yields (see below for a more
detailed discussion). Motivated by this we do not consider
lithium-7. In the CMB and $\text{D}+^{4}\text{He}$ concordance ranges
the standard BBN scenario overproduces $^7\text{Li}$
\cite{10.1093/ptep/ptac097}. In a faster expanding Universe results
are therefore worsen.  With the results from
Sec.~\ref{sec:standard_case} we then proceed to analyze the different
non-standard scenarios in Sec.~\ref{sec:nonstandard_case}.

Among the light elements abundances those of deuterium and
$^4\text{He}$ are the better measured. The ideal proxy for primordial
deuterium measurements are chemically unprocessed environments, with
the most precise determination following from a combination of 16
different measurements from high-redshift and low-metallicity quasar
absorption systems using large telescopes \cite{10.1093/ptep/ptac097}:
\begin{equation}
  \label{eq:d_abundance}
  \text{D}/\text{H}|_\text{p}=(25.36\pm 0.26)\times 10^{-6}\ .
\end{equation}
Helium-4 measurements, instead, follow from observations of
recombination emission lines of hydrogen and helium in blue compact
galaxies (typically at low redshifts). The wealth of data from these
sources has allowed a percent determination of the primordial baryon
fraction in $^4\text{He}$. The latest measurement obtained by
combining data from the Sloan Digital Sky Survey (SDSS), blue compact
dwarf galaxies \cite{Izotov:2003xn,Izotov:2007rq} and the Primordial
Helium Legacy Experiment with Keck (PHLEK) \cite{Hsyu:2020uqb} is:
\begin{equation}
  \label{eq:he4}
  \text{Y}_\text{p}=0.2436^{+0.0039}_{-0.0040}\ .
\end{equation}
Given the precision with which these two quantities are measured, any
approach to BBN aiming at reliable predictions should account for
higher-order effects. However, a semi-quantitative understanding
involving only leading effects is desirable as it captures the main
features of the intricate physics taking place during the BBN
era. Moreover, it enables a detailed understanding of the effects
implied by ultrastiff fluids. This is what
Sec.~\ref{sec:standard_case} aims at.

\begin{figure*}
  \centering
  \includegraphics[scale=0.37]{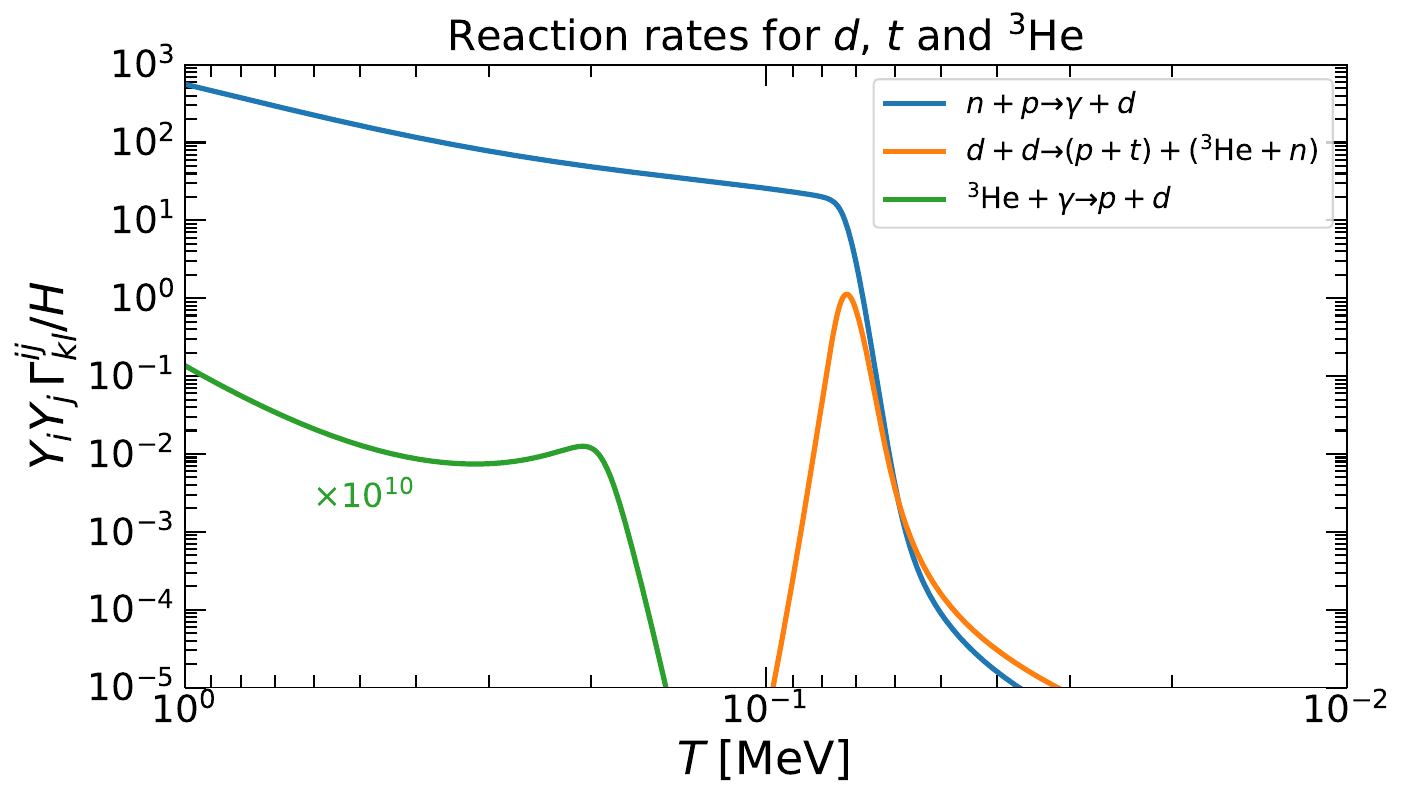}
  \includegraphics[scale=0.37]{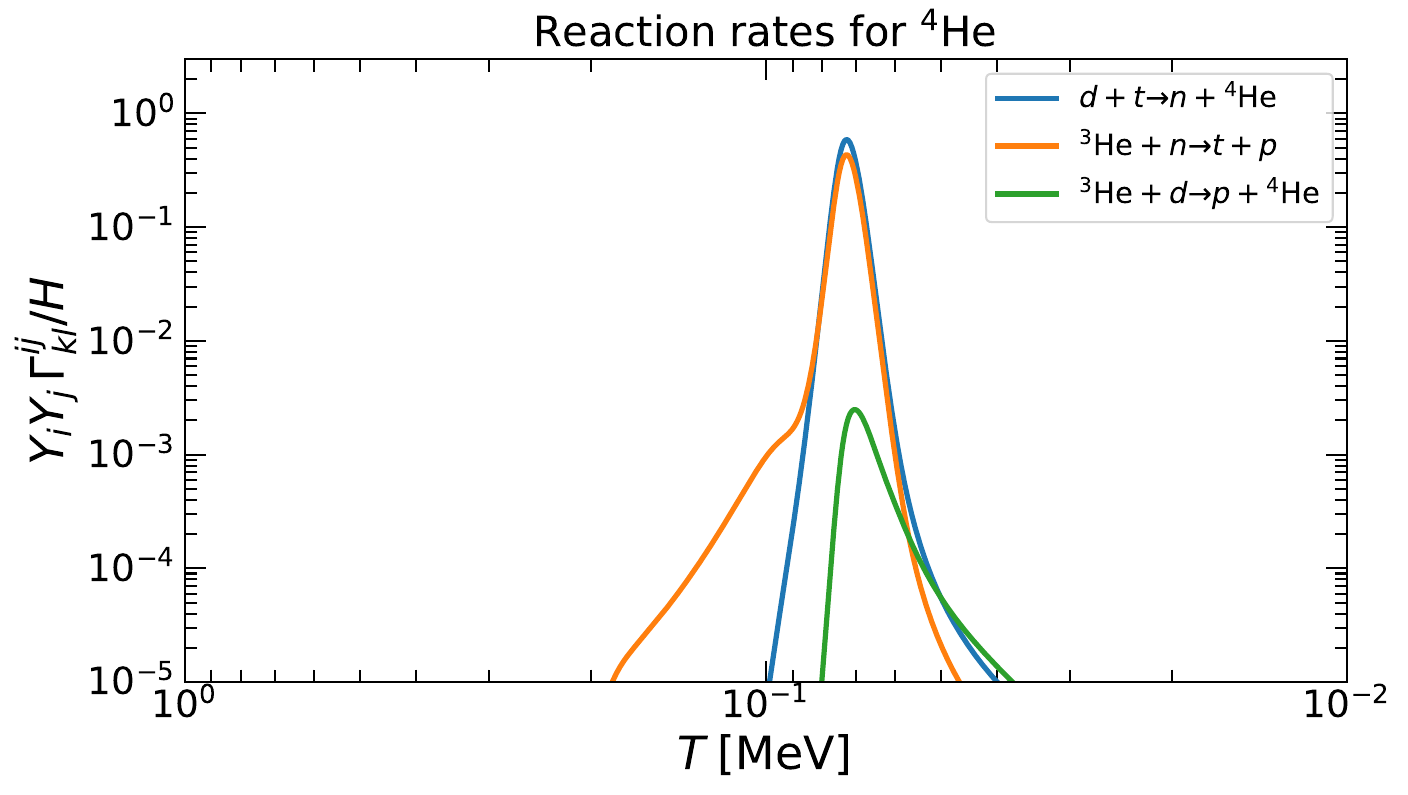}
  \includegraphics[scale=0.37]{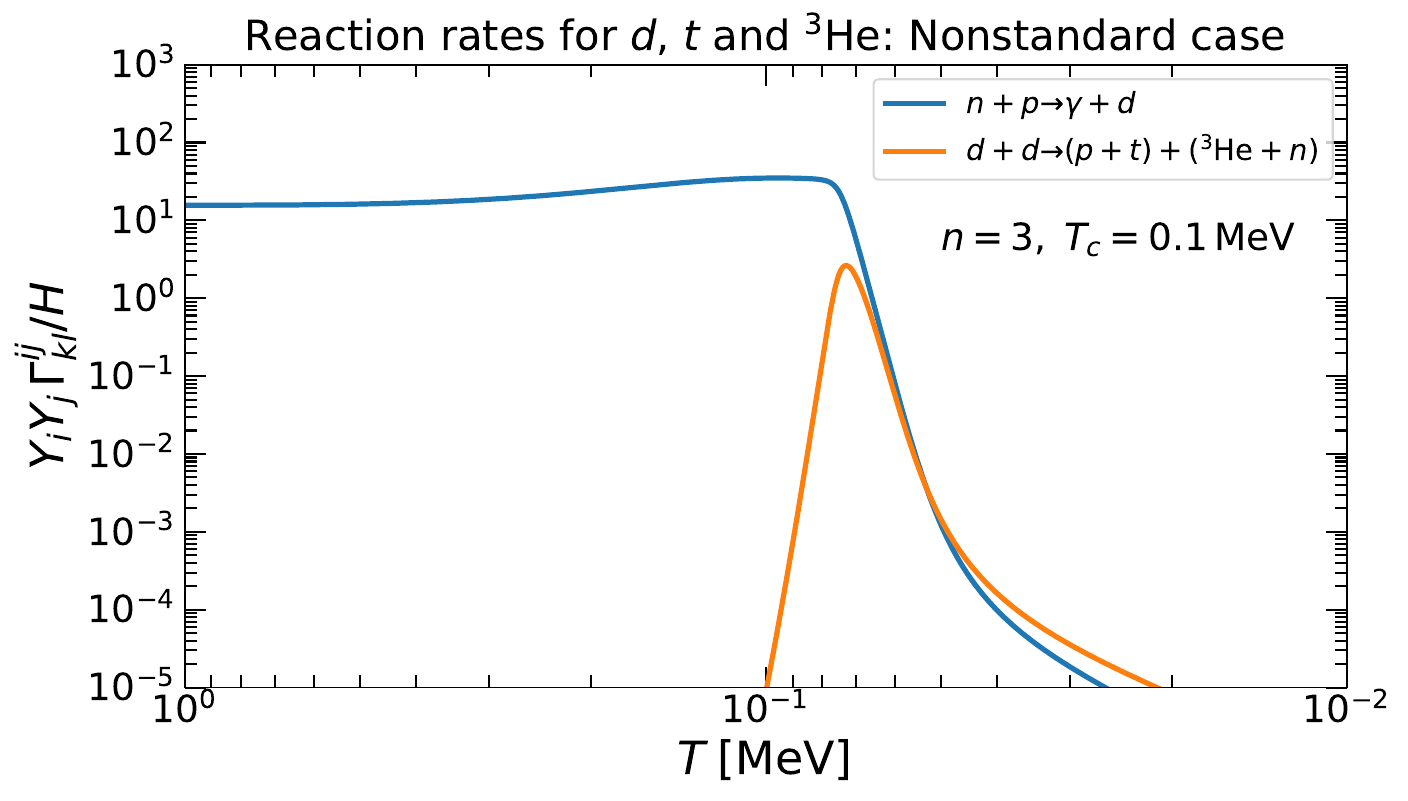}
  \includegraphics[scale=0.37]{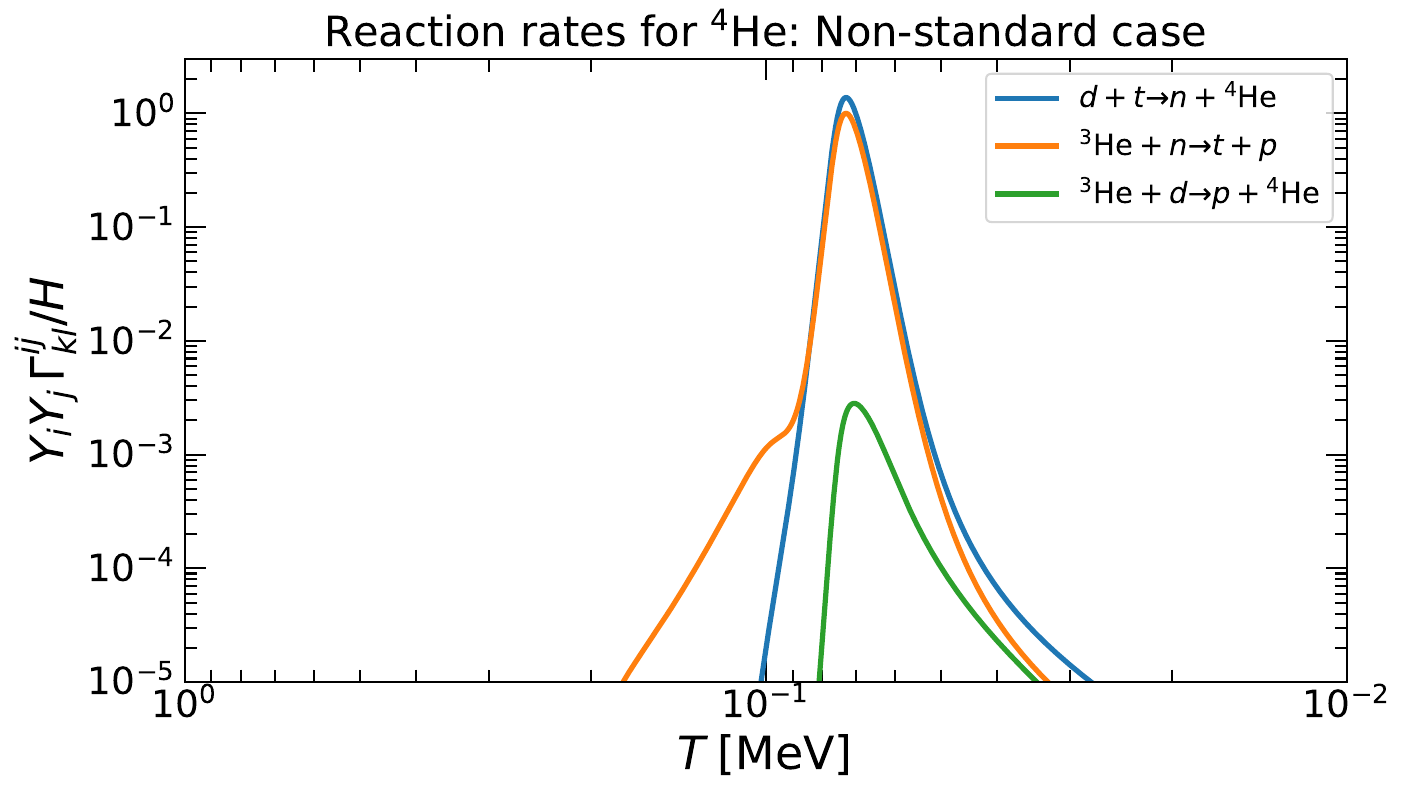}
  \caption{\textbf{Top left}: Most relevant reactions for the
    generation of the deuterium, tritium and $^{3}\text{He}$
    abundances in the standard scenario. Helium-3 photon dissociation
    is shown only for comparison. \textbf{Top right}: Most relevant
    processes for the production of $^{4}\text{He}$ in the standard
    scenario. \textbf{Bottom left}: Same as for the top left graph but
    for the non-standard case with an ultrastiff background with $n=3$
    and crossover temperature $T_c=0.1\,$MeV. \textbf{Bottom right}:
    Same as for the top right graph but for the non-standard case with
    an ultrastiff background with $n=3$ and crossover temperature
    $T_c=0.1\,$MeV.
    $\Gamma^{ij}_{kl}\equiv \rho_BN_A\langle\sigma\,v\rangle_{ij\to
      kl}$, where $\langle\sigma\,v\rangle_{ij\to kl}$ refers to the
    thermally-averaged cross section for the process $i+j\to k+l$ and
    $N_A$ is the Avogadro number. Explicit expressions for this
    quantity for the different reactions displayed in the graph are
    taken from \cite{Smith:1992yy}.}
  \label{fig:he4_plots}
\end{figure*}
Results presented in this section were obtained with the aid of
\texttt{PArthENoPE} and a few were double-checked with \texttt{PRIMAT}
\cite{Pisanti:2007hk,Consiglio:2017pot,Gariazzo:2021iiu,Pitrou:2018cgg,Pitrou:2020etk}.
\subsection{Building the light elements abundances: Standard scenario}
\label{sec:standard_case}
\begin{figure*}[t]
  \centering
  \includegraphics[scale=0.45]{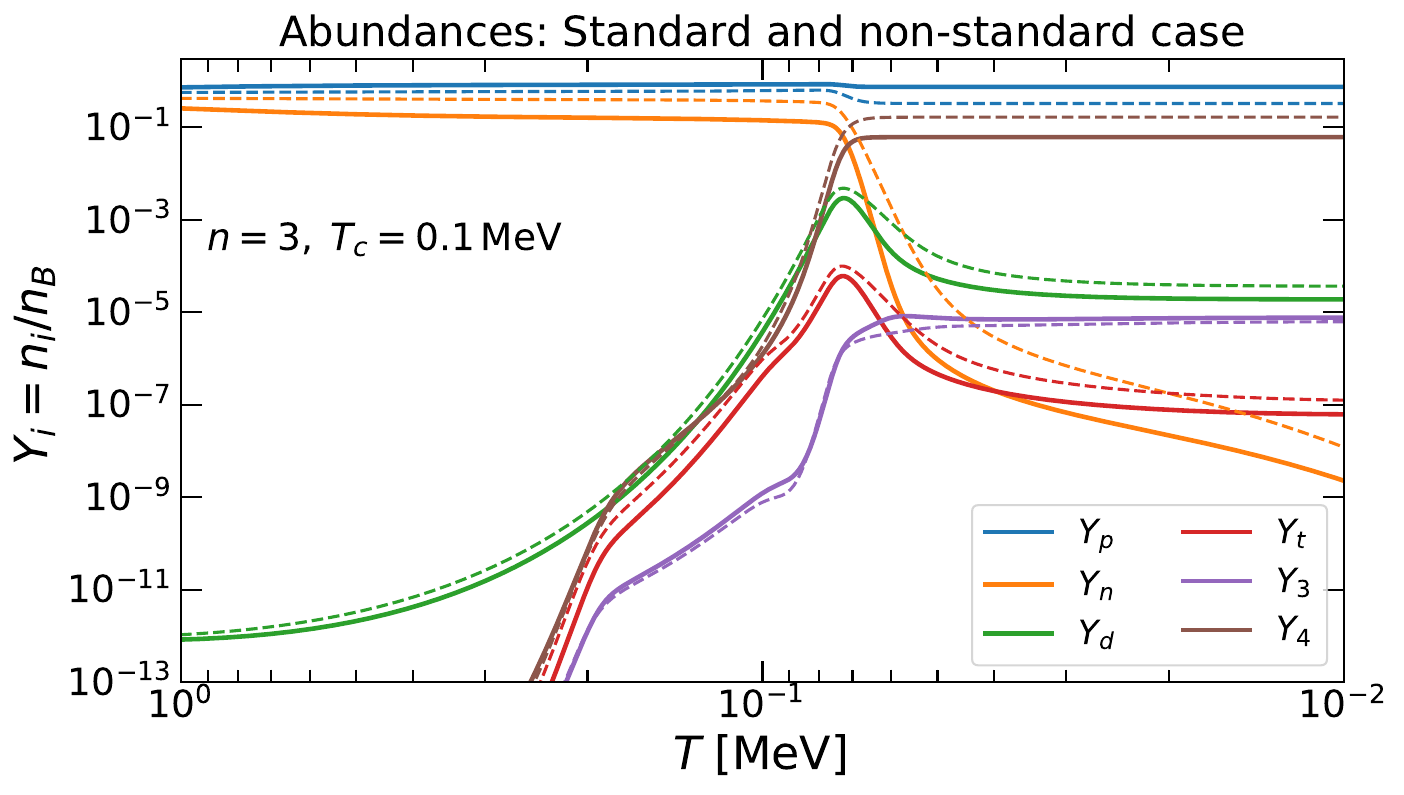}
  \caption{Evolution of the light elements abundances in the standard
    scenario as well as in the non-standard scenario $n=3$ and
    $T_c=0.1\,$MeV (dashed lines). Results have been derived with
    \texttt{PArthENoPE}
    \cite{Pisanti:2007hk,Consiglio:2017pot,Gariazzo:2021iiu} properly
    modified to account for deviations from the standard case. Results
    follow from $\Omega_B\,h^2=0.023$ and the neutron lifetime fixed
    to $\tau_n=885.7\,$sec. The latter inline with neutron beam
    experiments (see text for details). $Y_3$ and $Y_4$ refer to the
    abundances of $^3\text{He}$ and $^4\text{He}$, respectively.}
  \label{fig:abundances}
\end{figure*}

At temperatures of the order of 10 MeV, neutrons and protons are in
thermal contact with the primordial plasma. As the temperature
decreases, weak interactions slow down and so the neutron-to-proton
ratio deviates from its thermal distribution as can be seen in
Fig.~\ref{fig:deuterium_plots}. The ratio is further suppressed
because of neutron decay, which ceases when neutrons become bounded in
nuclei. At temperatures of the order of 1 MeV, the most efficient
nuclear process is
\begin{equation}
  \label{eq:pn_to_dgamma}
  p + n \leftrightarrow d + \gamma\ .
\end{equation}
The rate for the production process normalized to the expansion rate
is displayed in the top-left graph in Fig.~\ref{fig:he4_plots}. For
comparison the helium-3 photon dissociation process
$^{3}\text{He}+\gamma\to p+d$---magnified by $10^{10}$---is displayed
as well. At these temperatures forward and backward reactions are
taking place at the same rate and so the deuterium abundance building
up at this stage is still tiny: Although the temperature of the heat
bath is below the deuterium binding energy (2.3 MeV), photon
dissociation is still effective enough. The amount of entropy in the
plasma is sufficiently large for the high-energy tail of the photon
distribution to still entail a large amount of high-energy photons,
capable of breaking down the deuterium binding. As soon as photon
dissociation becomes ineffective---at
$T\simeq 7\times 10^{-2}\,$MeV---a rapid growth of the deuterium
abundance is then triggered (the deuterium abundance reaches its
peak), until the deuterium washout processes
\begin{equation}
  \label{eq:deuterium_washout}
  d+d\to p+t\quad\text{and}\quad d+d\to ^{3}\text{He}+n
\end{equation}
kick in at $T\simeq 6\times 10^{-2}\,$MeV (see the top-left graph in
Fig.~\ref{fig:he4_plots}). As the temperature decreases below this
value, these processes washout the deuterium abundance from its peak
at $\text{D}/\text{H}\simeq 3.65\times 10^{-3}$ down to its freeze-out
value $\text{D}/\text{H}=2.51\times 10^{-5}$ (results for
$Y_i=n_i/n_B$ are shown in Fig.~\ref{fig:abundances}).

The same processes that washout the deuterium abundance, source the
formation of tritium and helium-3. So, as the deuterium abundance
grows so does the tritium and helium-3 abundances. And this triggers
the growth of helium-4 in the plasma through three (leading)
processes:
\begin{align}
  \label{eq:he4_t}
  t+d&\to ^4\text{He}+n\ ,
  \\
  \label{eq:he4_he3}
  ^3\text{He}+d&\to ^4\text{He}+p\ ,
  \\
  \label{eq:he3_t}
  ^3\text{He}+n&\to t+p\ .
\end{align}
As can be seen in the top-right graph in Fig.~\ref{fig:he4_plots},
production through tritium annihilation is rather efficient up to
$T\simeq 7\times 10^{-2}\,$MeV. Direct production through
$^3\text{He}$ is instead subdominant, but $^{3}\text{He}-n$
annihilation sources tritium which then is reprocessed into
$^4\text{He}$. The neutrons left from the process in
Eq.~(\ref{eq:he4_t}) enhance the reaction rate. At lower
temperatures---of the order of $6\times 10^{-2}\,$MeV---process
(\ref{eq:he4_he3}) becomes relevant and contributes to $^4\text{He}$,
although at a much lower rate. Right after $^4\text{He}$ freezes out a
large fraction of neutrons are gone, as they have fueled its
formation.

In summary, tritium and $^3\text{He}$ source $^4\text{He}$ formation,
and so their abundance is smaller. Generation of tritium and
$^3\text{He}$ is sourced by deuterium, which in turn is sourced by
neutrons and protons. Along the way deuterium and $^3\text{He}$
washout processes inject protons back into the plasma, but neutrons
instead are depleted because ultimately they source all the abundances
and more importantly $^4\text{He}$ formation. The amount of neutrons
at the onset of BBN thus critically determines its output. The
evolution of the different abundances is shown in
Fig.~\ref{fig:abundances}.

\subsection{Building the light elements abundances in non-standard
  scenarios}
\label{sec:nonstandard_case}
As can be seen in Fig.~\ref{fig:deuterium_plots}, the
neutron-to-proton ratio (number of neutrons and protons available at
the onset of BBN) strongly depends on the type of ultrastiff fluid
present in that era and on $T_c$. Lower values of crossover
temperature can readily generate, for a sufficiently stiff
contribution, a freeze-out neutron-to-proton ratio that exceeds the
standard value by more than $75\%$. With such deviations, and
according to the discussion in the previous section, this enhancement
can have dramatic consequences on BBN predictions.

\begin{figure*}[t]
  \centering
  \includegraphics[scale=0.36]{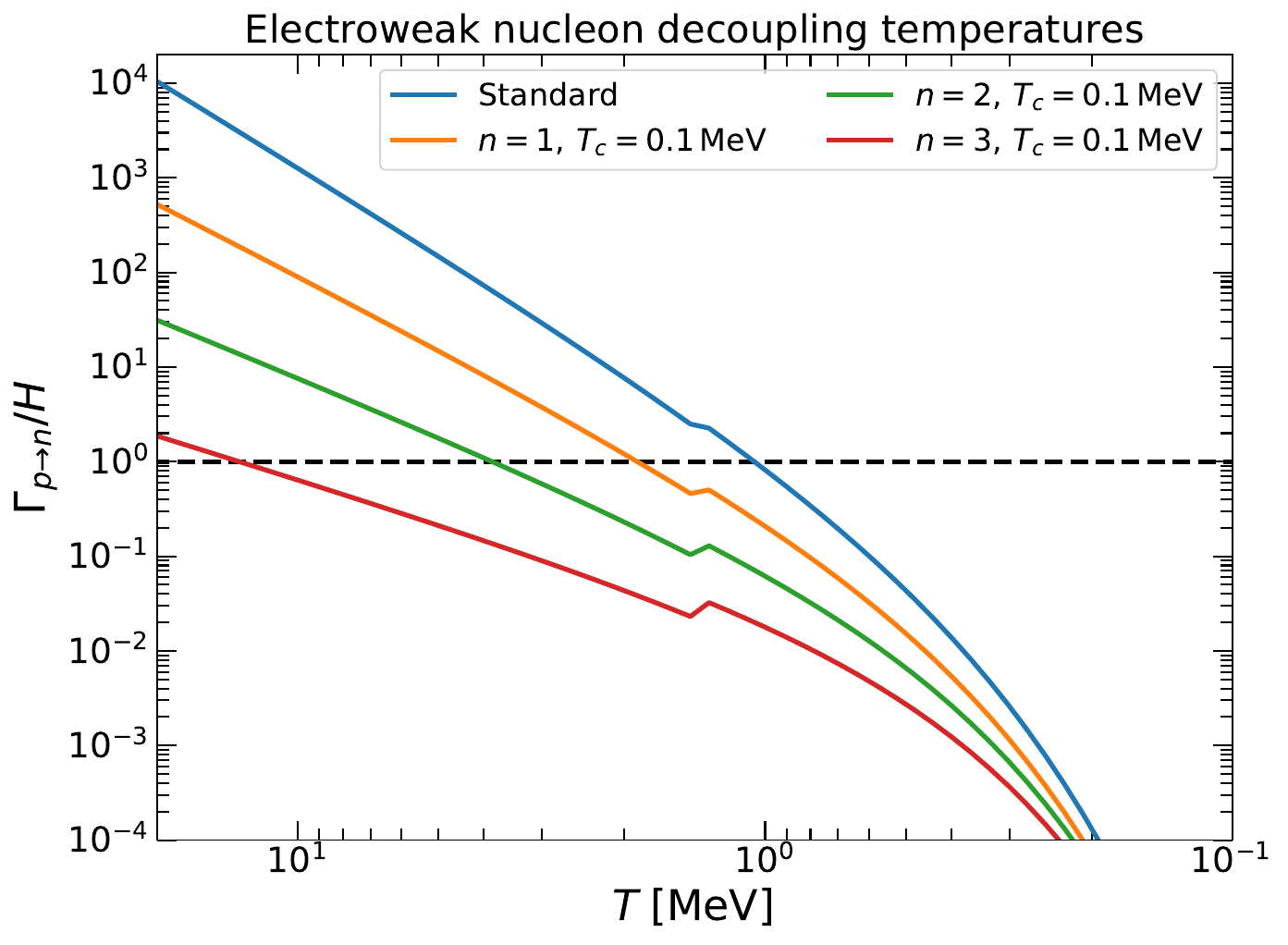}
  \caption{Leading-order proton electroweak reaction rate
    $\Gamma_{p\to
      n}\equiv\Gamma_{p+e\to\bar\nu_e+n}+\Gamma_{p+\nu_e\to\bar e+n}$
    normalized to the expansion rate in the standard case as well as
    in three different non-standard cases, with $T_c=0.1\,$MeV used as
    a proxy. It can be seen that non-standard cosmic expansion
    histories lead to nucleon decoupling at higher temperature.}
  \label{fig:proton_rate_versus_H}
\end{figure*}

\begin{figure*}[t]
  \centering
  \includegraphics[scale=0.375]{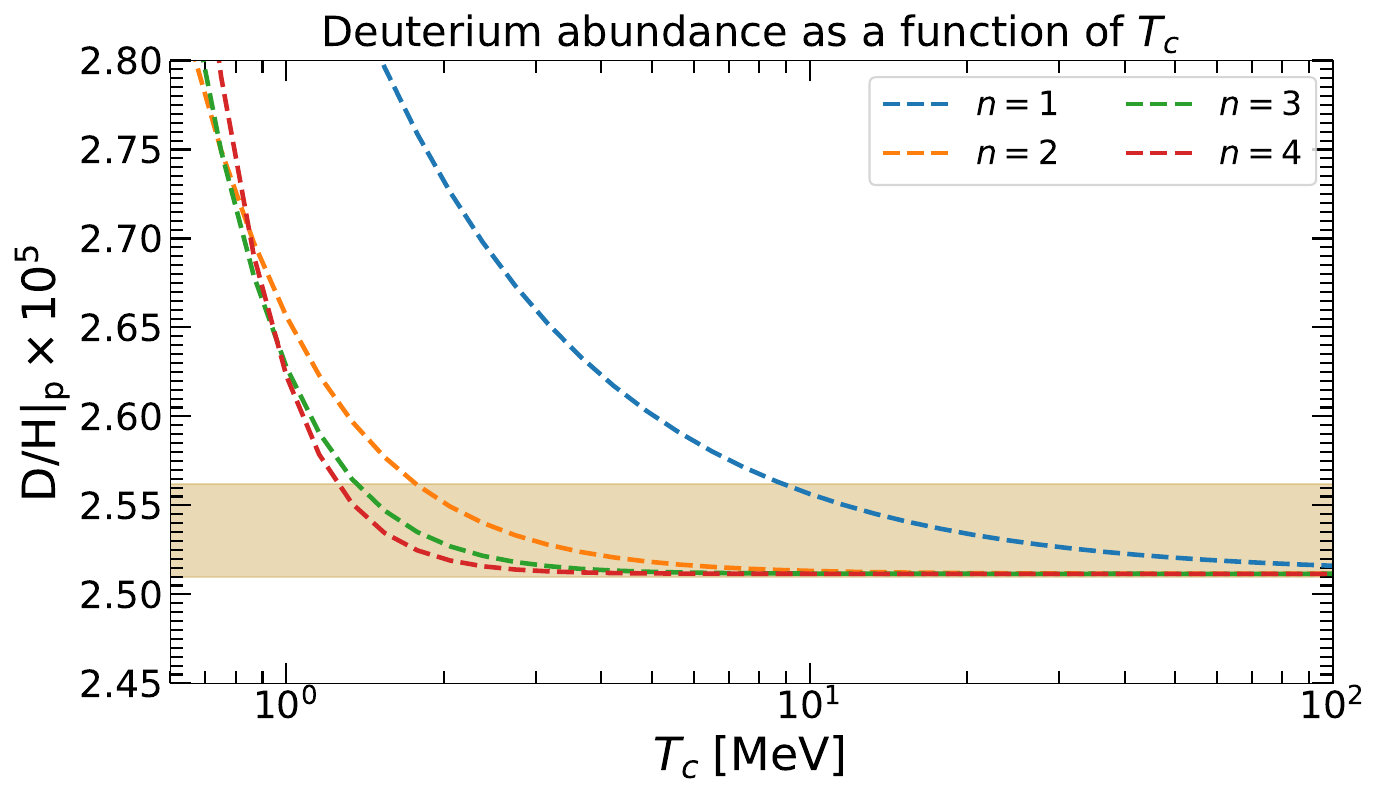}
  \includegraphics[scale=0.375]{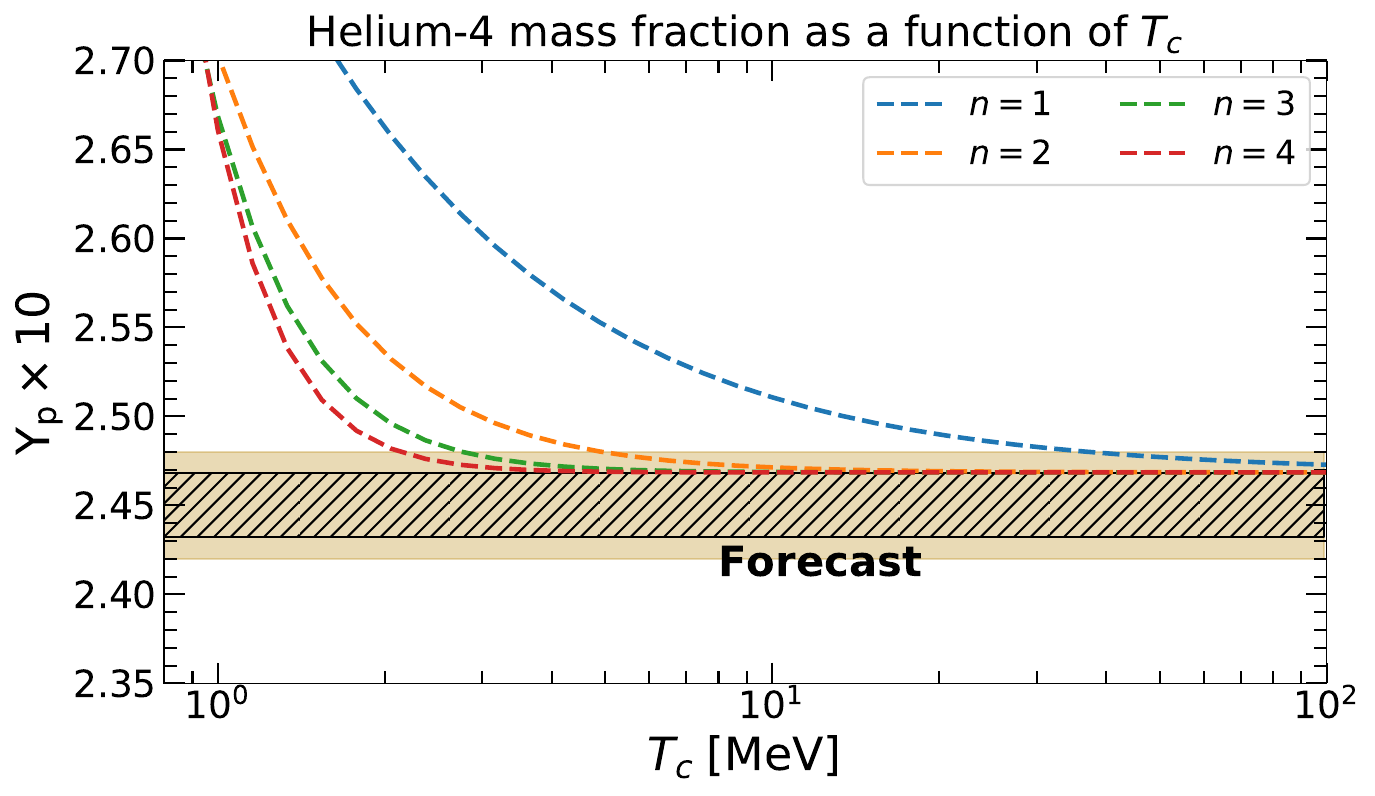}
  \includegraphics[scale=0.375]{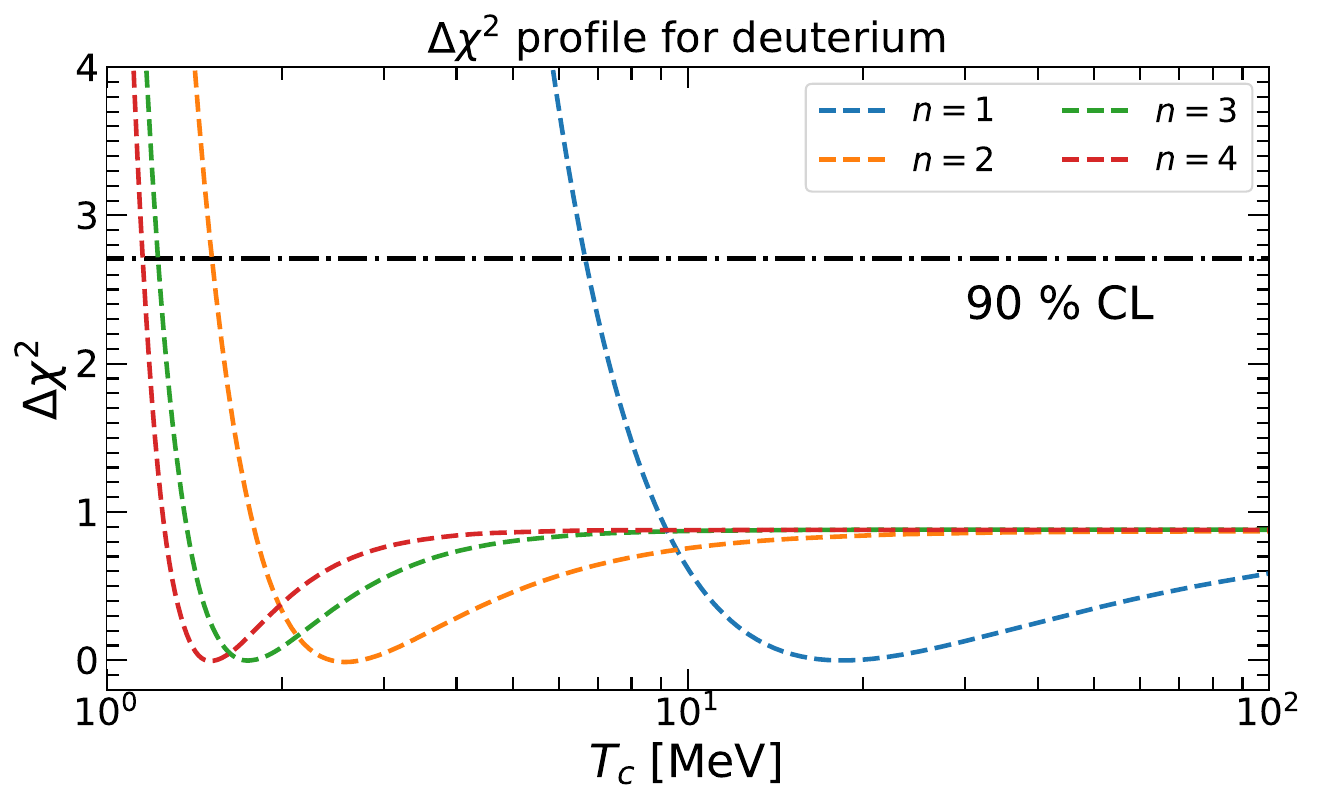}
  \includegraphics[scale=0.375]{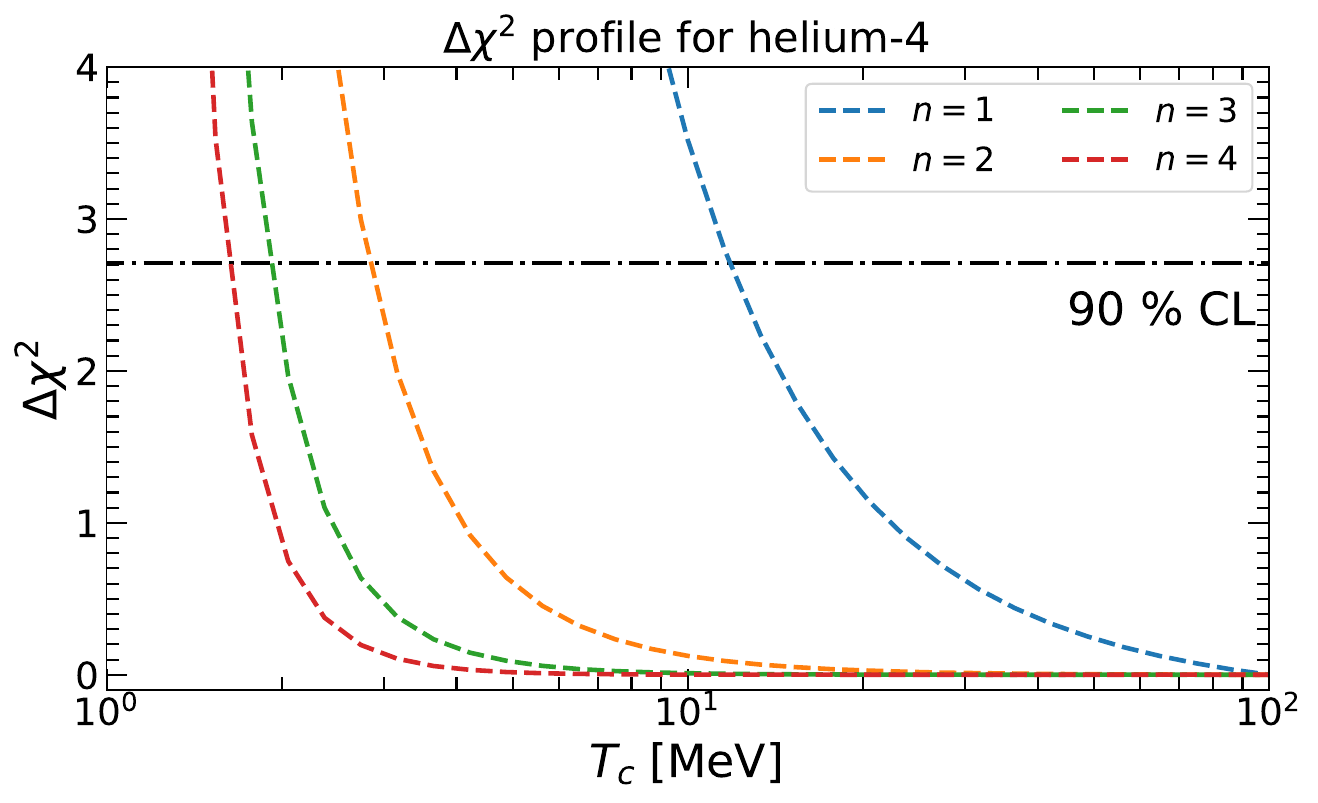}
  \caption{\textbf{Top}: Deuterium (left graph) and helium-4 (right
    graph) abundances $\text{D}/\text{H}|_\text{p}$ and
    $\text{Y}_\text{p}$ as a function of crossover temperature $T_c$
    for exponential indices $n=1,\cdots, 4$. The stripes correspond to
    allowed measured intervals given in Eqs.~(\ref{eq:d_abundance})
    and (\ref{eq:he4}) and taken from
    Refs.~\cite{10.1093/ptep/ptac097,Hsyu:2020uqb}. The hatched region
    correspond to a forecast where the statistical uncertainty has
    been required to close the region where ultrastiff fluids can
    generate a sizable deviation from the standard
    expectation. \textbf{Bottom}: $\Delta\chi^2$ profiles for
    deuterium (left graph) and helium (right graph) as a function of
    $T_c$. The $90\%$ CL limit on $T_c$ can be directly read from both
    graphs. The absence of an absolute minimum for $\Delta\chi^2$ in
    the helium-4 case is due to the theoretical prediction never
    reaching the measured central value.}
  \label{fig:deuterium_and_he4_abundances}
\end{figure*}

A larger neutron-to-proton ratio is expected because of the faster
expansion rate. Let us discuss this in more detail. Coupled with the
ultrastiff fluid, the modified expansion rate is steeper than the
standard one and so electroweak interactions decouple at higher $T$
\footnote{Usually, weak freeze-out is quoted to happen at
  $T\simeq 0.8\;$MeV. However, a proper inclusion of lepton capture on
  neutrons and protons as well as blocking factors for in-medium free
  neutron decay has demonstrated that they are still active close to
  the epoch of $^{4}\text{He}$ formation ($T\simeq 100\,$keV)
  \cite{Grohs:2016vef}.}. Equivalently, electroweak interactions
happen less often because of the faster expansion and so they undergo
freeze-out at earlier times. This is illustrated in
Fig.~\ref{fig:proton_rate_versus_H}, where we have plotted the
leading-order proton reaction rate
$\Gamma_{p\to n}\equiv
\Gamma_{p+e\to\bar\nu_e+n}+\Gamma_{p+\nu_e\to\bar e+n}$ (see Appendix
\ref{sec:proton_neutron_rates} for explicit expressions) normalized to
the expansion rate as a function of $T$.  Instantaneous neutrino
decoupling has been assumed at $T=1.4\,$MeV. Note that this result is
displayed only to illustrate the earlier decoupling induced by the
non-standard expansion. Results presented in
Figs.~\ref{fig:abundances} along with those in
Fig.~\ref{fig:deuterium_and_he4_abundances} involve all electroweak
decoupling effects included in \texttt{PArthENoPE}.

The most remarkable consequence of electroweak decoupling at higher
$T$ is that of increasing the helium-4 abundance. The baryon fraction
in helium-4 quoted in Eq.~(\ref{eq:he4}) can be well approximated by
\begin{equation}
  \label{eq:baryon_fraction_in_he4}
  \text{Y}_\text{p}=\frac{2\,n_n/n_p}{1 + n_n/n_p}\ ,
\end{equation}
from which it becomes clear that a substantial modification of the
neutron-to-proton ratio will, in turn, drastically modify its BBN
prediction. Note that in the absence of the ultrastiff fluid, standard
case, the $n_n/n_p$ freeze-out temperature is controlled by the
neutron lifetime. A larger $\tau_n$ means electroweak interactions are
weaken, resulting in an earlier decoupling and so in an overproduction
of $^{4}\text{He}$.

Motivated by a $3.6\,\sigma$ mismatch between neutron beam and
ultra-cold neutron (UCN) storage methods measurements
\cite{Yue:2013qrc,Pattie:2017vsj}
\begin{align}
  \label{eq:neutron_lifetime}
  \tau_n^\text{Beam}&=887.2\pm 2.2\,\text{sec}\ ,
  \\
  \tau_n^\text{UCN}&=877.7\pm 0.7\,\text{sec}\ ,
\end{align}
the lifetime of the neutron has been recently a matter of
debate. Indeed, the impact of such mismatch has motivated the study of
its implications in BBN predictions \cite{Chowdhury:2022ahn}. Here it
is worth stressing that our results demonstrate that deviations on the
$^4\text{He}$ abundance---that could be attributed to $\tau_n$---can
be as well interpreted/understood in terms of a primordial era
dominated by an ultrastiff fluid. The size of the deviation determined
by the nature of the fluid (exponent $n$) and the crossover
temperature.

That $^{4}\text{He}$ is overproduced can be seen directly in
Fig.~\ref{fig:abundances}, where we have calculated, along with the
standard case, results for the case $n=3$ and $T_c=0.1\,$MeV. This
choice, although just a particular case in parameter space, shows the
impact of a sufficiently steep energy density contribution. At
freeze-out, the standard calculation leads to
$\text{Y}_\text{p}=0.2469$ while the prediction with an expansion
determined by the chosen parameters produces
$\text{Y}_\text{p}=0.6693$, a number clearly ruled out by data [see
Eq.~(\ref{eq:he4})]. From the same Figure one can as well see that
deuterium is also largely overproduced,
$\text{D}/\text{H}|_\text{p}=1.1\times 10^{-4}$. The number density
increases by about a factor of two and the hydrogen (protons)
abundance decreases by about the same amount, thus resulting in a
$\sim 75\%$ deviation compared with the standard result and ruled out
by data [see Eq.~(\ref{eq:d_abundance})]. Tritium follows the same
trend, its value increases sizably compared to the standard
result. Unlike these elements, helium-3, however, is slightly
depleted.

The behavior just described in the previous paragraph can be
understood by relying on the leading-order processes governing BBN,
and discussed in Sec.~\ref{sec:standard_case} for the standard
case. Comparing the top and bottom left graphs in
Fig.~\ref{fig:he4_plots} one can see that at the onset of BBN
($T\simeq 0.1\,$MeV) deuterium production is more efficient in the
non-standard case. This is the case because of the initial-state
nucleon density being larger by about a factor of 2. From those graphs
one can also note that the processes that populate helium-3 and
tritium abundances become effective at slightly lower $T$, but still
enough for the deuterium abundance to reach a higher peak and to
decrease less abruptly (see Fig.~\ref{fig:abundances}). Since
deuterium is more abundant, tritium is more efficiently produced and
so its abundance is enhanced. By the same token one would expect the
helium-3 abundance to be enhanced, but this is not the case. The
reason has to do with the process in Eq.~(\ref{eq:he3_t}). Since
neutrons are more abundant, the $^3\text{He}-n$ annihilation rate is
higher, resulting in a depletion of its abundance. Note that unlike
the standard case $^3\text{He}-d$ annihilation---that will otherwise
contribute to deuterium and further helium-3 depletion---is
inefficient all across the relevant temperature range.

With the general trend already understood, we are left with the
determination of the minimal crossover temperature ($T_c^\text{min}$)
for which the ultrastiff fluid is still consistent with current
$\text{D}/\text{H}|_\text{p}$ and $\text{Y}_\text{p}$ measurements
[see Eqs.~(\ref{eq:d_abundance}) and (\ref{eq:he4})]. For that aim we
have calculated these abundances as a function of $T_c$ for different
ultrastiff fluid scenarios, $n=1,\cdots, 4$. Results are shown in
Fig.~\ref{fig:deuterium_and_he4_abundances}, top-left (top-right)
graph for deuterium ($^{4}\text{He}$). Inline with our previous
discussion, $^4\text{He}$ places tighter limits on $T_c$ (regardless
of the ultrastiff fluid nature). The temperature behavior, however,
matches in both cases. The less stiff the background is, the higher
the $T_c$ for which consistency with data is found. This is readily
understood from the temperature dependence of the ultrastiff fluid
energy density [see Eq.~(\ref{eq:rho_stiff})]. Because of the boundary
condition, ultrastiff-radiation equality is fixed universally
regardless of $n$. At temperatures above $T_c$, larger values of $n$
produce a broader $\rho_s$ domination epoch. In contrast, below $T_c$
ultrastiff components fade away more abruptly with $T$. Thus, for
small values of $n$ ultrstiff-radiation equality must happen at larger
$T$ to prevent the background component to contribute sizably.

Simple inspection of the results indicate that consistency with
deuterium data requires $T_c\gtrsim \{8.8,1.8,1.2,1.0\}\,$MeV for
$n=\{1,2,3,4\}$. For $^4\text{He}$, instead, results are
$T_c\gtrsim \{32.0,5.0,3.0,2.0\}\,$MeV for $n$ covering the same
values. To put these results on a more statistical basis we have
performed a $\chi^2$ analysis with the aid of the following
least-square function
\begin{equation}
  \label{eq:chi2_BBN}
  \chi^2_\text{BBN}(T_c,n)=
  \left[
    \frac{\mathcal{A}_\text{Th}(T_c,n) - \mathcal{A}_\text{Exp}}
    {\sigma_\mathcal{A}}
  \right]^2\ ,
\end{equation}
where $\mathcal{A}$ stands for deuterium or $^4\text{He}$ abundances
and $\sigma_\mathcal{A}$ for the statistical uncertainty in
Eqs.~(\ref{eq:d_abundance}) and (\ref{eq:he4}). Results are shown in
Fig.~\ref{fig:deuterium_and_he4_abundances} for deuterium (helium-4)
in the bottom-left (bottom-right) graph. Values for $T_c$ at the
$90\%$ CL are listed in Tab.~\ref{tab:TC_BBN}. As already mentioned,
$^4\text{He}$ places the tightest limits, although at this level values
are rather comparable to those obtained with deuterium data. Note that
values listed in Tab.~\ref{tab:TC_BBN} are rather inline with those
derived in Ref.~\cite{DEramo:2017gpl}, for arbitrary $n$. They differ
by only $20\%-30\%$.
\begin{table}[h!]
  \centering
  \setlength{\tabcolsep}{7pt}
  \begin{tabular}{|c|c|c|c|c|}\hline
    $n$  & 1 & 2 & 3 & 4 \\\hline\hline
    $T_c^\text{min}\;\;(\text{D}/\text{H}|_\text{p}) [\text{MeV}]$
         & 6.660 & 1.518 & 1.226 & 1.154\\\hline
    $T_c^\text{min}\;\;(\text{Y}_\text{p}) [\text{MeV}]$
         & 11.80 & 2.838 & 1.903 & 1.613 \\\hline
  \end{tabular}
  \caption{$90\%$ CL lower limit for $T_c$ and for cases
    $n=1,\cdots,4$. The results follow after calculating the relative
    abundance for deuterium and $^4\text{He}$ under the presence of
    the ultrastiff fluid component and with the aid of the $\chi^2$
    function in Eq.~(\ref{eq:chi2_BBN}).}
  \label{tab:TC_BBN}
\end{table}

In addition to current primordial helium-4 measurements---that imply
values in Tab.~\ref{tab:TC_BBN}---we have considered as well a
forecast. To do so, we have set the condition that the statistical
uncertainty of such hypothetical measurement allows fully closing the
region where the ultrastiff fluid generates a sizable departure from
the standard expectation. We have found that a measurement reducing
the statistical uncertainty by $\sim 60\%$ will test all the relevant
values of $T_c$ for which a departure from expectation is observable,
at least for the first four realizations $n=1,\cdots,4$
($\omega=2/3,1,4/3,5/3$) of the ``full tower''. The hatched region in
the top-right graph in Fig.~\ref{fig:deuterium_and_he4_abundances}
shows the result.

\section{Conclusions}
\label{sec:conclusions}%
In this paper we have considered the impact of ultrastiff fluids in
early Universe dynamics, which emerge in cosmological scenarios with a
non-standard equation of state. Since their energy density can
dominate at early times they can---potentially---affect the Universe
expansion rate and contribute sizably to the amount of ``radiation''
during that epoch. Given that BBN data provides the most early picture
of the early Universe, we have focused our analysis on the possible
traces left by this type of backgrounds on neutrino decoupling and BBN
observables.

Considering neutrino decoupling seems---to a certain
extent---mandatory. First of all, is during this process that the
photon distribution temperature gets overheated (above that of the
neutrinos). And in a faster expanding Universe that overheating is
expected to be affected, mainly because of the impact on the
temperature where electron-positron pair annihilation ceases. An
overheating of the photon distribution above standard expectations has
consequences on $N_\text{eff}$, beyond those already implied by the
ultrastiff fluid.

A faster expansion rate is also expected to affect the network of
kinetic equations accounting for the formation of the light elements
abundances. From a very simplified point of view, the mere fact that
the neutron-to-proton ratio can be dramatically changed at the onset
of BBN already suggest that light elements abundances are good
``tracers'' of the imprints left by these fluids.

Our findings are as follows. Neutrino decoupling alone, coupled with
indirect measurements of $N_\text{eff}$ inferred from CMB power
spectra, are able to test some regions in the $n-T_c$ plane. Their
extent, however, is mild and so allow for crossover temperatures to be
as small as $10^{-1}\,$MeV depending on stiffness. Improvements in
$N_\text{eff}$ measurements, as those expected in stage-III or
stage-IV CMB experiments, can change that conclusion. If they were to
be considered alone, expectations from these forthcoming measurements
could improve sensitivities by a factor of a few.

When BBN data is used instead, sensitivities from $N_\text{eff}$ are
proven to be within regions that are hard to reconcile with
measurements of primordial helium-4 abundances. Thus, a deviation of
$N_\text{eff}$ from standard expectations cannot be accounted for by
this type of early Universe physics. Primordial helium-4 and deuterium
measurements, in particular the former, have the largest
sensitivities. Less stiff scenarios are more prone to the constraints
implied by these data. For $n=1$, ultrastiff-radiation equality needs
to happen above $\sim 12\,$MeV, i.e. much before the onset of neutrino
decoupling and BBN. With increasing stiffness ultrastiff-radiation
equality can happen at lower temperatures. In the case of stiff matter
($\omega=1$, $n=2$) that equality can be as low as $3\,$MeV, while
still being consistent with current primordial helium-4 measurements.

We have shown that improving the statistical uncertainty in the
primordial helium-4 abundance measurement by $\sim 60\%$, will
increase sensitivities to regions where still these fluids can leave
traceable tracks. So, with such improvement on data they can be fully
tested. Beyond those sensitivities, the regions that can be explored
produce signals that degenerate with standard expectations and so
sensitivities are lost. If after enhancing the statistical uncertainty
to this value no deviations are observed, neither neutrino decoupling
nor BBN observables will be suited tools for testing this hypothesis.
\appendix
\section{Relevant thermodynamic quantities and collision terms}
\label{sec:thermodynamics}
In this appendix we collect the expressions required to run the
network of Boltzmann equations that account for the neutrino
decoupling process, Eqs.~(\ref{eqs:Beqs_nu}) and
(\ref{eqs:Beqs_gam}). We start with standard results for energy
density and pressure of relativistic thermal species
\cite{Alpher:1953zz}
\begin{align}
  \label{eq:density_pressure}
  \rho_\gamma=\frac{\pi^2}{30}g_\gamma T^4\ ,\quad\rho_\nu=\frac{7}{8}\rho_\gamma\ ,
\end{align}
with $g_\gamma=g_\nu=2$. Pressure follows from these expressions:
$p_i=\rho_i/3$ ($i=\gamma,\nu$). At the time of $e^+e^-$ annihilation,
electron and positrons undergo a transition from the relativistic to
non-relativistic regimes. Energy density as well as pressure has then
to be written in the general form \cite{Alpher:1953zz}
\begin{align}
  \label{eq:density_e}
  \rho_e&= \frac{2g_e}{2\pi^2}T^4\int_{\overline{m}_e}^\infty
          \frac{\left(x^2-\overline{m}_e^2\right)^{1/2}}
          {e^x + 1}x^2\,dx\ ,
  \\
  \label{eq:pressure_e}
  p_e&=\frac{2g_e}{6\pi^2}T^4\int_{\overline{m}_e}^\infty
       \frac{\left(x^2-\overline{m}^2_e\right)^{3/2}}
       {e^x + 1}\,dx\ ,
\end{align}
where the electron chemical potential has been set to zero,
$g_e=g_{e^-}=g_{e^+}=2$ and $\overline{m}_e\equiv m_e/T$. Photon and
neutrino temperature partial derivatives follow directly from
Eqs.~(\ref{eq:density_pressure}) and are given by
$\partial\rho_i/\partial T_i=4\rho_i/T_i$. For electrons, instead, by
\begin{equation}
  \label{eq:partial_derivatives}
  \frac{\partial}{\partial T} \rho_e=\frac{2\cdot 4\,g_e}{2\pi^2}
  T^3\int_{\overline{m}_e}^\infty
  \frac{\left(x^2-\overline{m}_e^2\right)^{1/2}}{\cosh^2(x/2)}x^3\,dx\ .
\end{equation}
Collision terms follow from electron, positron, neutrino and
anti-neutrino electroweak neutral and charged current scattering
processes (contributions from $s$, $t$ and $u$ channels are present)
\cite{Dolgov:2002wy}. Assuming a Maxwell-Boltzmann distribution and
$T_{\nu_i}\equiv T_\nu$, following Ref.~\cite{Escudero:2018mvt} they
can be written according to
\begin{align}
  \label{eq:collision_terms_nue}
  \frac{\delta\rho_{\nu_e}}{\delta t}&=\frac{G_F^2}{\pi^5}
                                   \left[
                                   (1 + 4\sin^2\theta_W
                                   + 8\sin^4\theta_W)F(T,T_\nu)
                                   \right]\ ,
  \\
  \label{eq:collision_terms_numu}
  \frac{\delta\rho_{\nu_\mu}}{\delta t}&=\frac{G_F^2}{\pi^5}
                                   \left[
                                   (1 - 4\sin^2\theta_W
                                   + 8\sin^4\theta_W)F(T,T_\nu)
                                   \right]\ ,
\end{align}
with the function $F$ given by \cite{Escudero:2018mvt}
\begin{equation}
  \label{eq:maxwell_boltzmann_function}
  F(T_1,T_2)=32(T_1^9-T_2^9)+56T_1^4T_2^4(T_1-T_2)\ .
\end{equation}
For the weak mixing angle we have employed $\sin^2\theta_W=0.223$
\cite{10.1093/ptep/ptac097}.
\section{Proton leading-order rate in the infinite mass nucleon limit}
\label{sec:proton_neutron_rates}
The leading-order proton reaction rate $\Gamma_{p\to n}$ consist of
two terms: $\Gamma_{p+e\to \bar\nu_e + n}$ and
$\Gamma_{p+\nu_e\to \bar e + n}$. Both can be written in terms of the
integral $\lambda_{p\to n}$ \cite{Alpher:1953zz,Weinberg:2008zzc}
\begin{widetext}
  \begin{equation}
    \label{eq:lambda_pn}
    \lambda_{p\to n}(q_\text{min},q_\text{max}) = \int_{q_\text{min}}^{q_\text{max}}
    \sqrt{1-\frac{m_e}{q+Q}}\left(q+Q\right)^2q^2f(q+Q,T)
    \left[1 - f(q,T_\nu)\right]\,dq\ ,
  \end{equation}
\end{widetext}
where $Q=m_n-m_p$, $E_e=q+Q$ and
$f(x,y)=(e^{x/y}+1)^{-1}$. Integration limits determine the process
for which $\lambda_{p\to n}$ accounts for. For $p+e\to \bar\nu_e + n$,
limits are $q_\text{min}=-\infty$ and $q_\text{max}=-(m_e+Q)$.  For
$p+\nu_e\to \bar e + n$, limits read $q_\text{min}=0$ and
$q_\text{max}=\infty$. In terms of Eq.~(\ref{eq:lambda_pn}) the proton
reaction rate is then given by
\begin{equation}
  \label{eq:Gamma_p_n}
  \Gamma_{p\to n} =
  \frac{\lambda_{p\to n}(-\infty,-m_e-Q) + \lambda_{p\to n}(0,\infty)}
  {m_e^5\tau_n\lambda_0}\ ,
\end{equation}
with the coefficient $\lambda_0$ written according to \cite{Pitrou:2018cgg}
\begin{equation}
  \label{eq:lambda0}
  \lambda_0 =\sqrt{\overline{\Delta}^2-1}
  \left(
    \frac{-8-9\overline{\Delta}^2+2\overline{\Delta}^4}{60}
  \right)
  +
  \frac{\overline\Delta}{4}\operatorname{arcosh}(\overline\Delta)\ ,
\end{equation}
where $\overline\Delta \equiv Q/m_e$.
\section*{Acknowledgments}
We warmly thank Nicol\'as Fern\'andez, Kazunori Kohri and Robert Scherrer for very useful comments on the
manuscript.  A.V.\ thanks the Theoretical Astroparticle and Neutrino
Physics Group of the University of Torino and INFN for their
hospitality. The work of D.A.S.\ and A.V.\ is supported by ANID under
project number 1221445.  S.G.\ thanks ``Universidad T{\'e}cnica
Federico Santa Mar{\'i}a'' in Santiago de Chile for hospitality in the
months during which this project was conceived and carried out.  S.G.\
is supported by the European Union’s Framework Programme for Research
and Innovation Horizon 2020 (2014–2020) under grant agreement 754496
(FELLINI, until September 2023) and Junior Leader Fellowship LCF/BQ/PI23/11970034 by La Caixa Foundation (from October 2023).
\bibliography{references_cosmo}

\end{document}